\documentclass[12pt]{article}
\usepackage{amssymb}
\usepackage{amsmath}

\begin{document}

\def\binom#1#2{{#1\choose #2}}

\title{On problem of polarization tomography, I}
%(first draft)}
\author{Roman Novikov
%\thanks{}
\\
CNRS,
Laboratoire de\\ Math\'{e}matiques Jean Leray,\\
Universit\'{e} de Nantes,\\
BP 92208, F-44322,\\
Nantes cedex 03, France\\
\begin{sloppypar}
novikov@math.univ-nantes.fr
\end{sloppypar}
\and
Vladimir Sharafutdinov\thanks{The second author was supported by
RFBR grant 05-01-00611 and DFG - RFBR grant
04-01-04000. The
work was started when the author visited Universit\'{e} de Nantes June
2006, and the work was mostly done when the author visited
Universit\"{a}t des Saarlandes in the scope of
the latter grant.
The author is grateful to the whole faculty of the
Applied Mathematics Department of the university
and especially to his landlady, Frau Reinhardt, for her
hospitality.}
\\
%
% and by RFBR Grant 05--01--00611.}\\
Sobolev Institute of Mathematics\\
4 Koptjug Avenue\\
Novosibirsk, 630090, Russia\\
\begin{sloppypar}
sharaf@math.nsc.ru
\end{sloppypar}
}
\date{}
%September 2006, Saarbruecken

\maketitle

\setcounter{section}{0}
\setcounter{page}{1}
\newtheorem{Theorem}{Theorem}[section]
\newtheorem{Lemma}[Theorem]{Lemma}
\newtheorem{Problem}[Theorem]{Problem}
\newtheorem{Corollary}[Theorem]{Corollary}
%\pagestyle{headings}
%\markboth{}{}
%\date{December 30, 1999}

\begin{abstract}
The polarization tomography problem consists of recovering a matrix
function $ f $ from the fundamental matrix of the equation
$ D\eta/dt=\pi_{\dot\gamma}f\eta $ known for every geodesic $ \gamma $
of a given Riemannian metric.
Here $ \pi_{\dot\gamma} $ is the orthogonal projection onto the hyperplane
$ \dot\gamma{}^\bot $.
The problem arises in optical tomography
of slightly anisotropic media. The local uniqueness theorem is proved:
a $ C^1 $-small function $ f $ can be recovered from the data uniquely
up to a natural obstruction. A partial global result is obtained
in the case of the Euclidean metric on $ {\mathbb R}^3 $.
\end{abstract}

%\noindent
%{\bf Key words:}
%Integral geometry, Ray transform, Pseudodifferential operators

%\noindent
%{\bf AMS Subject Classification:}

\section[]{Introduction}

First of all we shortly recall the physical motivation of the problem.
See Section 5.1 of \cite{[mb]} for a detailed discussion.

We consider propagation of time-harmonic electromagnetic waves of frequency
$ \omega $ in a medium with the zero conductivity, unit magnetic
permeability, and the dielectric permeability tensor of the form
$$
\varepsilon_{ij}=n^2\delta_{ij}+
\frac {1} {k}\chi_{ij},
                        \eqno{(1.1)}
$$
where $ k=\omega/c $ is the wave number, $ c $ being the light
velocity. Here $ n>0 $ is a function of a point $ x\in{\mathbb R}^3 $,
and the tensor $ \chi_{ij}=\chi_{ij}(x) $ determines a small anisotropy
of the medium. The smallness is emphasized by the factor $ 1/k $.
Equation (1.1) was suggested by Yu.~Kravtsov \cite{[K]}. By some
physical arguments \cite{[KO]}, the tensor $ \chi $ must be Hermitian,
$ \chi_{ij}={\bar\chi}_{ji} $.

In the scope of the zero approximation of geometric optics, propagation
of electromagnetic waves in such media is described as follows. Exactly
as in the background isotropic medium, light rays are geodesics of the
Riemannian metric
$$
dt^2=n^2(x)|dx|^2;
                        \eqno{(1.2)}
$$
the electric vector $ E(x) $ and magnetic vector $ H(x) $ are
orthogonal to each other as well as to the ray; and the amplitude
$ A^2=|E|^2=|H|^2 $ satisfies $ A=C/\sqrt{nJ} $ along a ray, where $ J
$ is the geometric divergence and the constant $ C $ depends on the ray.
The only difference between a slightly anisotropic medium and the
background isotropic one consists of the wave polarization. The
polarization vector $ \eta=n^{-1}A^{-1}E $ satisfies the equation
(generalized Rytov's law)
$$
\frac {D\eta} {dt}=\frac {i} {2n^2}\pi_{\dot\gamma}\chi\eta
                        \eqno{(1.3)}
$$
along a geodesic ray $ \gamma(t) $. Here
$ t $ is the arc length of $\gamma$ in metric (1.2),
$ \dot\gamma=d\gamma/dt $ is the speed
vector of $ \gamma $, $ \pi_{\dot\gamma} $ is the orthogonal projection
onto the plane $ \dot\gamma{}^\bot $, and
$ D/dt=\dot\gamma{}^k{\nabla}_{\!k} $ is the covariant derivative along
$ \gamma $ in metric (1.2). The right-hand side of (1.3) is understood
as follows: $ \pi_{\dot\gamma} $ and $ \chi $ are considered as linear
operators on $ T^{\mathbb C}_{\gamma(t)}={\mathbb C}^3 $,
and $ \pi_{\dot\gamma}\chi\eta $ is
the result of action of the operator $ \pi_{\dot\gamma}\chi $ on the
complex vector $ \eta\in\dot\gamma{}^\bot $. Here $ \dot\gamma{}^\bot $
is the two-dimensional complex vector space consisting of complex
vectors orthogonal to the real vector
$ \dot\gamma\in{\mathbb R}^3=T_{\gamma(t)}\subset
T^{\mathbb C}_{\gamma(t)}={\mathbb C}^3  $. Introducing the
notation
$$
f=\frac {i} {2n^2}\chi,
                        \eqno{(1.4)}
$$
we rewrite (1.3) in the form
$$
\frac {D\eta} {dt}=\pi_{\dot\gamma}f\eta.
                        \eqno{(1.5)}
$$
Observe that $ f $
is a skew-Hermitian operator, $ f_{ij}=-{\bar f}_{ji}
$.

Let us now consider the inverse problem. Assume a medium under
investigation to be contained in a bounded domain
$ D\subset{\mathbb R}^3 $ with a smooth boundary. The background
isotropic medium is assumed to be known, i.e., metric (1.2) is given.
The domain $ D $ is assumed to be convex with respect to the metric,
i.e., for any two boundary points $ x_0,x_1\in\partial D $, there
exists a unique geodesic $ \gamma:[0,1]\rightarrow D $ such that
$ \gamma(0)=x_0,\ \gamma(1)=x_1 $. We consider the inverse problem of
determining the anisotropic part $ \chi_{ij} $ of the dielectric
permeability tensor or, equivalently, of
determining the tensor $ f $ on (1.5). To
this end we can fulfill tomographic measurements of the following type.
For any unit speed geodesic $ \gamma:[0,l]\rightarrow D $ between
boundary points, we can choose an initial value
$ \eta_0=\eta(0)\in\dot\gamma{}^\bot(0) $ of the polarization vector
and measure the final value $ \eta_1=\eta(l)\in\dot\gamma{}^\bot(l) $
of the solution to equation (1.5). In other words, we assume the linear
operator
$ \dot\gamma{}^\bot(0)\rightarrow\dot\gamma{}^\bot(l),\
\eta_0\mapsto\eta_1 $ to be known for every unit speed geodesic
$ \gamma:[0,l]\rightarrow D $ between boundary points. Instead of
(1.5), we will consider the corresponding operator equation
$$
\frac {D\tilde U(t)} {dt}=
f_{\dot\gamma(t)}\tilde U(t),
                        \eqno{(1.6)}
$$
where $ f_{\dot\gamma(t)}:\dot\gamma{}^\bot(t)
\rightarrow\dot\gamma{}^\bot(t) $ is the restriction of the operator
$ \pi_{\dot\gamma(t)}f(\gamma(t)) $ to the plane
$ \dot\gamma{}^\bot(t) $, and the solution is considered as a linear
operator
$ \tilde U(t):\dot\gamma{}^\bot(t)\rightarrow\dot\gamma{}^\bot(t) $.
Equation (1.6) has a unique solution satisfying the initial condition
$$
\tilde U(0)=E,
                        \eqno{(1.7)}
$$
where $ E $ is the identity operator. Since $ f_{\dot\gamma(t)} $ is a
skew-Hermitian operator, the solution $ \tilde U(t) $ is a unitary
operator. The final value of the solution
$$
\tilde\Phi[f](\gamma)=\tilde U(l)\in GL(\dot\gamma{}^\bot(l))
$$
is the data for
the inverse problem.  Given the function
$ \tilde\Phi[f] $ on the set of unit speed geodesics
between boundary points, we have
to determine the tensor field $ f=(f_{ij}(x)) $ on the domain $ D $.

We consider the inverse problem in a more general setting. Instead
of a domain $ D\subset{\mathbb R}^3 $ with metric (1.2), we will
consider a compact Riemannian manifold $ (M,g) $ of an arbitrary
dimension $ n\geq 3 $, and an arbitrary complex tensor field
$ f=(f_{ij}) $ on $ M $. In such a setting, equation (1.6) makes sense
along a geodesic $ \gamma $. We will subordinate the manifold $ (M,g) $
to some conditions that guarantee smoothness of the data
$ \tilde\Phi[f] $ in the case of a smooth $ f $.

The two-dimensional case of $ n=2 $ is not
interesting since $ f_{\dot\gamma} $ and $ \tilde U $ become scalar
functions and the solution to the scalar equation (1.6) is given by an
explicit formula in this case. Therefore the inverse problem is reduced
to the inversion of the ray transform $ I $ on second rank tensor
fields, see the remark before Theorem 5.2.1 of \cite{[mb]}.

Equation (1.6) can be slightly simplified. For a point $ x\in M $, let
$ T^{\mathbb C}_xM $ be the complexification of the tangent space
$ T_xM $. Instead of considering the operator $ \tilde U(t) $ on
$ \dot\gamma{}^\bot(t) $, we define the linear operator
$$
U(t):T^{\mathbb C}_{\gamma(t)}M\rightarrow T^{\mathbb C}_{\gamma(t)}M
$$
by
$$
U(t)|_{\dot\gamma{}^\bot(t)}=\tilde U(t),\quad
U(t)\dot\gamma(t)=\dot\gamma(t).
$$
If $ \tilde U(t) $ satisfies (1.6)--(1.7), then $ U(t) $ solves the
initial value problem
$$
\frac {DU} {dt}=(\pi_{\dot\gamma}f\pi_{\dot\gamma})U,
\quad U(0)=E.
                        \eqno{(1.8)}
$$
Equation (1.8) is more handy than (1.6) since all operators
participating in (1.8) are defined on the whole of
$ T^{\mathbb C}_{\gamma(t)}M $. The inverse problem consists of
recovering the tensor field $ f $ from the data
$ \Phi[f](\gamma)=U(l) $ known for every unit speed geodesic
$ \gamma:[0,l]\rightarrow M $ between boundary points.
Let us emphasize that the inverse problem is
strongly nonlinear, i.e., the data $ \Phi [f] $ depends on $ f $ in a
nonlinear manner.

The three-dimensional case, $ n=\mbox{dim}\,M=3 $, is of the most
importance for applications as we have shown above. On the other hand,
the three-dimensional case is mathematically the exceptional one
because, for a skew-symmetric $ f $, the solution to the inverse
problem is not unique. The non-uniqueness is discussed in Section 4.

The main result of the present article is the local uniqueness theorem:
the solution to the inverse problem is unique (up to a natural
obstruction in the three-dimensional case) if the tensor field $ f $ is
$ C^1 $-small. See Theorem 5.1 below for the precise statement.

Our method of investigating the inverse problem is a combination of
approaches used in \cite{[V]} and in Chapter 5 of \cite{[mb]}. First of
all, following \cite{[V]}, we reduce our nonlinear problem to a linear
one as follows. Let $ f_i\ (i=1,2) $ be two tensor fields and $ U_i(t)
$ be the corresponding solutions to (1.8) with $ f=f_i $. Then
$ u=U^{-1}_1U_2-E $ satisfies
$$
\frac {Du} {dt}=p\pi_{\dot\gamma}(f_2-f_1)\pi_{\dot\gamma}q,
\quad u(0)=0,
                        \eqno{(1.9)}
$$
where $ p=U^{-1}_1 $ and $ q=U_2 $. We consider $ p $ and $ q $ as
operator-valued weights which are close to the unit operator if $ f_i $
are $ C^1 $-small. Assuming the weights $ p $ and $ q $ to be fixed,
the solution $ u(t) $ to the initial value problem (1.9) depends
linearly on $ f=f_2-f_1 $.
We study the linear inverse problem of recovering the tensor field
$ f=f_2-f_1 $ from the data $ F[f](\gamma)=u(l) $ given for all
geodesics $ \gamma:[0,l]\rightarrow M $ between boundary points.
In the case of a symmetric tensor field $ f
$ and of unit weights, this linear problem was considered in Chapter 5
of \cite{[mb]}. We will demonstrate that the same approach works in the
case of an arbitrary $ f $ and of weights close to the unit one.

There is one more opportunity to extract a
linear inverse problem from equation (1.8). Indeed, if
$ W(t)=\mbox{det}\,U(t) $ is the Wronskian, then the function
$ \varphi(t)=\ln W(t) $ satisfies
$$
\frac {d\varphi(t)} {dt}=\mbox{tr}\,(\pi_{\dot\gamma}
f\pi_{\dot\gamma}).
$$
Therefore, for every unit speed geodesic
$ \gamma:[0,l]\rightarrow M $ between boundary points, the integral
$$
S[f](\gamma)=
\int\limits_{0}^{l}\mbox{tr}\,(\pi_{\dot\gamma(t)}
f(\gamma(t))\pi_{\dot\gamma(t)})dt
                        \eqno{(1.10)}
$$
is expressed through the data $ \Phi[f] $ by the formula
$$
S[f](\gamma)=\ln\mbox{det}\,\Phi[f](\gamma).
$$
The data $ S[f] $ depends linearly on $ f $. Of course, some
information is lost while the data $ \Phi[f] $ is replaced with
$ S[f] $. In particular, $ S[f] $ is independent of the skew-symmetric
part of $ f $.

We finally note that main results of the article are new and
nontrivial in the case of $ M\subset{\mathbb R}^n $, $ n\geq 3 $, with
the standard Euclidean
metric. If a reader is not familiar with the tensor analysis machinery
on the tangent bundle of a Riemannian manifold, he/she can first read
the article for the latter simplest case.

The article is organized as follows. Section 2 contains some
preliminaries concerning Riemannian geometry and tensor analysis. In
particular, we define some class of Riemannian manifolds for which the
problem can be posed in the most natural way. Instead of
considering the ordinary differential equation (1.8) along individual
geodesics, we introduce a partial differential equation on the unit
tangent bundle and pose an equivalent version of the problem in terms
of the latter equation. In Section 3, we consider the corresponding
linear problem and prove the uniqueness for weights sufficiently close
to the unit in the case of $ n\geq 4 $. Section 4 discusses the
three-dimensional case. In Section 5, we check that the weights
$ p $ and $ q $ are sufficiently close to the unit for a $ C^1 $-small
$ f $ and prove our main result, Theorem 5.1, on the local uniqueness in
the nonlinear problem. In the final Section 6, we investigate the
question: to which extent is a symmetric tensor field $ f $ determined
by data (1.10). We give a complete answer to the question in the case
of $ M={\mathbb R}^3 $ with the Euclidean metric.

%The author is grateful to R. Novikov for a useful discussion of the
%subject. In particular, R.~Novikov introduced the operator $ S $
%defined by (1.10).

\section[]{Posing the problem and introducing\\
some notations}

A smooth compact Riemannian manifold $ (M,g) $
with boundary is said to be a
{\it convex non-trapping manifold} (CNTM briefly) if it
satisfies two conditions:
(1) the boundary $ \partial M $ is strictly convex, i.e., the second
fundamental form
$$
\mbox{II}(\xi,\xi)=\langle {\nabla}_{\!\xi}\nu,\xi\rangle
\quad \mbox{for}\quad \xi\in T_x(\partial M)
$$
is positive definite for every boundary point $ x\in\partial M $, where
$ \nu $ is the outward unit normal vector to the boundary
and $ {\nabla}_{\!\xi} $ is the covariant derivative in the
direction $ \xi $; and
(2) for every $ x\in M $ and $ 0\neq\xi\in T_xM $, the maximal geodesic
$ \gamma_{x,\xi}(t) $
determined by the initial conditions $ \gamma_{x,\xi}(0)=x $
and $ \dot\gamma_{x,\xi}(0)=\xi $ is defined on a finite segment
$ [\tau_-(x,\xi),\tau_+(x,\xi)] $. In what follows, we use the
notations $ \gamma_{x,\xi} $ and $ \tau_\pm(x,\xi) $ many times. They
are always understood in the sense of this definition.

{\bf Remark.} In \cite{[mb]}, the term CDRM (compact dissipative
Riemannian manifold) is used instead of CNTM. In the case of
$ M\subset{\mathbb R}^n $ with the standard Euclidean metric, this
definition means that $ M $ is strictly convex.

By $ TM=\{(x,\xi)\mid x\in M, \xi\in T_xM\} $ we denote the tangent
bundle and by
$$
\Omega M=\{(x,\xi)\in TM\mid |\xi|^2=\langle\xi,\xi\rangle=
g_{ij}(x)\xi^i\xi^j=1\},
$$
the unit sphere bundle. Its boundary can be represented as the union
$ \partial\Omega M=\partial_+\Omega M\cup\partial_-\Omega M $, where
$$
\partial_\pm\Omega M=
\{(x,\xi)\in\Omega M\mid x\in\partial M,\
\pm\langle\xi,\nu(x)\rangle\geq 0\}
$$
is the manifold of outward (inward) unit vectors. If
$ \gamma:[0,l]\rightarrow M $ is a unit speed geodesic
between boundary points, then
$ (\gamma(0),\dot\gamma(0))\in\partial_-\Omega M $ and
$ (\gamma(l),\dot\gamma(l))\in\partial_+\Omega M $.

By $ T_x^{\mathbb C}M $ we denote the complexification of the tangent
space $ T_xM $. The metric $ g $ determines the Hermitian scalar
product on $ T_x^{\mathbb C}M $
$$
\langle \eta,\zeta\rangle =g_{ij}\eta^i\bar\zeta{}^j.
                        \eqno{(2.1)}
$$
For $ 0\neq\xi\in T_xM $, by
$ T^\bot_{x,\xi}M=\{\eta\in T_x^{\mathbb C}M\mid\langle\eta,\xi\rangle=0\} $
we denote the orthogonal complement of $ \xi $ and by
$ \pi_\xi:T_x^{\mathbb C}M\rightarrow T_x^{\mathbb C}M $, the orthogonal
projection onto $ T^\bot_{x,\xi}M $.

Let $ \tau^r_sM $ be the bundle of complex tensors that are $ r $ times
contravariant an $ s $ times covariant. Elements of the section space
$ C^\infty(\tau ^r_sM) $ are smooth tensor fields of rank $ (r,s) $ on
$ M $. In the domain of a local coordinate system, such a field
$ u\in C^\infty(\tau ^r_sM) $ can be represented by the family of
smooth functions,
$ u=(u^{i_1\dots i_r}_{j_1\dots j_s}(x)) $, the coordinates of $ u $,
where each index takes values from 1 to $ n=\mbox{dim}\,M $.
The metric $ g $ determines canonical isomorphisms
$ \tau ^r_sM\cong\tau^{r+s}_0M\cong\tau^0_{r+s}M $. We will consider
the isomorphisms as identifications. So, we do not distinct contra- and
covariant tensors but use contra- and covariant coordinates of the same
tensor. For example, for $ u\in C^\infty(\tau^0_2M)=
C^\infty(\tau^1_1M)=C^\infty(\tau^2_0M) $,
$$
u_{ij}=g_{ik}u_{\cdot j}^{k\cdot}=g_{jk}u_{i\cdot}^{\cdot k}=
g_{ik}g_{jl}u^{kl}.
$$
In particular, such a tensor field determines the linear operator
$$
u(x):T^{\mathbb C}_xM\rightarrow T^{\mathbb C}_xM,\quad
(u\eta)^i=u^{i\cdot}_{\cdot j}\eta^j
$$
at any point $ x\in M $. The product of two such operators is written
in coordinates as $ (uv)_{ij}=u_{ik}v^{k\cdot}_{\cdot j} $. The dual
operator has the coordinates $ u^*_{ij}={\bar u}_{ji} $. The operator
is Hermitian (symmetric) if and only if $ u_{ij}={\bar u}_{ji} $
($ u_{ij}=u_{ji} $). The scalar product (2.1) is extended to tensors by
the formula
$ \langle u,v\rangle=u^{i_1\dots i_m}{\bar v}_{i_1\dots i_r} $ and
determines the norm $ |u|^2=\langle u,u\rangle $. For
$ u,v\in C^\infty(\tau^1_1M) $, the norm of the product satisfies
$ |uv|\leq\sqrt{n}|u||v| $, where $ n=\mbox{dim}\,M $.

We will also widely use semibasic tensor fields introduced in
\cite{[PS]}, see either Section 3.4 of \cite{[mb]} or Section 2.5 of
\cite{[ml]} for a detailed presentation.  Let $ \beta^r_sM $ be the
bundle of complex semibasic tensor fields of rank $ (r,s) $. It is a
subbundle of $ \tau^r_s(TM) $ isomorphic to
the induced bundle $ \pi^*(\tau^r_s M) $,
where $ \pi:TM\rightarrow M $ is the projection of the tangent
bundle.
A tensor
$ u\in T^r_{s,(x,\xi)}(TM) $ at $ (x,\xi)\in TM $ is semibasic if it is
``pure contravariant in the $ \xi $-variable and pure covariant in
$ x $'', i.e.,
$$
u=u^{i_1\dots i_r}_{j_1\dots j_s}
\frac {\partial} {\partial\xi^{i_1}}\otimes\dots\otimes
\frac {\partial} {\partial\xi^{i_r}}\otimes
dx^{j_1}\otimes\dots\otimes dx^{j_s}.
$$
For $ U\subset TM $, by $ C^\infty(\beta^r_sM;U) $ we denote
the space of smooth sections over $ U $. The notation $
C^\infty(\beta^r_sM;TM) $ is abbreviated to $ C^\infty(\beta ^r_sM)
$. In the domain of a local coordinate system, such a field $ u\in
C^\infty(\beta^r_sM) $ can be represented by the family of its
coordinates, $ u=(u^{i_1\dots i_r}_{j_1\dots j_s}(x,\xi)) $, which
are smooth functions of $ 2n $ variables $
(x,\xi)=(x^1,\dots,x^n,\xi^1,\dots,\xi^n) $. All the content of the
previous paragraph is extended to semibasic tensor fields,
where $ g $ remains the metric on $ M $ in the identification
of contra- and covariant tensors. In
particular, $ u\in C^\infty(\beta^1_1M) $ determines the linear
operator $ u(x,\xi):T^{\mathbb C}_xM\rightarrow T^{\mathbb C}_xM $
for every $ (x,\xi)\in TM $. There are two important first order
differential operators
$$
\stackrel v{\nabla},\stackrel h{\nabla}:
C^\infty(\beta^r_sM)\rightarrow C^\infty(\beta^r_{s+1}M)
$$
which are called the vertical and horizontal covariant derivatives.
The operators are defined in local coordinates by the formulas
$$
{\stackrel v{\nabla}}_{\!k}
u^{i_1\dots i_r}_{j_1\dots j_s}
=
\frac{\partial}{\partial\xi^k}
u^{i_1\dots i_r}_{j_1\dots j_s},
\quad\quad\quad\quad\quad\quad\quad\quad\quad\quad\quad
\quad\quad\quad
$$
$$
{\stackrel h{\nabla}}_{\!k}
u^{i_1\dots i_r}_{j_1\dots j_s}
=
\frac{\partial}{\partial x^k}
u^{i_1\dots i_r}_{j_1\dots j_s}-
\Gamma^p_{kq}\xi^q
\frac{\partial}{\partial\xi^p}
u^{i_1\dots i_r}_{j_1\dots j_s}
+
\quad\quad\quad\quad\quad\quad
$$
$$
\quad\quad\quad\quad\quad\quad
+
\sum\limits_{a=1}^{r}
\Gamma^{i_a}_{kp}
u^{i_1\dots i_{a-1}pi_{a+1}\dots i_r}_{j_1\dots j_s}
-
\sum\limits_{a=1}^{s}
\Gamma^{p}_{kj_a}
u^{i_1\dots i_r}_{j_1\dots j_{a-1}pj_{a+1}\dots j_s},
$$
where $\Gamma^i_{jk}$ are Christoffel symbols.
See
Sections 3.4--3.6 of \cite{[mb]} for properties of
these operators.
Note that $ \stackrel v{\nabla}=\partial/\partial\xi $ and
$ \stackrel h{\nabla}=\partial/\partial x $ in the case of
$ M\subset{\mathbb R}^n $ with the standard Euclidean metric and of
Cartesian coordinates.

The operator
$$
H=\xi^i{\stackrel h{\nabla}}_{\!i}:
C^\infty(\beta^r_sM)\rightarrow C^\infty(\beta^r_{s}M)
$$
is of the most importance in the present article. It is called the
differentiation with respect to the geodesic flow.

\bigskip

Given a tensor field $ f\in C^\infty(\tau^1_1M) $ on a CNTM
$ (M,g) $, let us consider the boundary value problem
$$
HU(x,\xi)=\pi_\xi f(x)\pi_\xi U(x,\xi)
\quad \mbox{on}\quad \Omega M,\quad \quad
U|_{\partial_-\Omega M}=E,
                        \eqno{(2.2)}
$$
where $ E $ is the identity operator. A solution $ U=U(x,\xi) $ is
assumed to be a section of the bundle $ \beta^1_1M $ over
$ \Omega M $, i.e., $ U\in C(\beta^1_1M;\Omega M) $.
In the case of $ M\subset{\mathbb R}^n $ with the standard Euclidean
metric, $ f $ and $ U $ can be considered as $ n\times n $-matrix
valued functions of $ x\in M $ and of
$ (x,\xi)\in M\times{\mathbb S}^{n-1} $ respectively.
Problem (2.2) has a
unique solution. Indeed, if we restrict (2.2) to an orbit of the
geodesic flow, i.e., if we set $ x=\gamma(t) $ and
$ \xi=\dot\gamma(t) $ for a unit speed geodesic
$ \gamma:[0,l]\rightarrow M $ with $ \gamma(0)\in\partial M $, then we
immediately arrive to the initial value problem (1.8). The boundary
value problem (2.2) is thus equivalent to the family of initial value
problems (1.8) considered for all unit speed geodesics simultaneously.
The inverse problem is now formulated as follows: one has to recover
the tensor field $ f $ given the trace
$$
\Phi[f]=U|_{\partial_+\Omega M}
                        \eqno{(2.3)}
$$
of the solution to (2.2).

In order to abbreviate further formulas, let us introduce the operator
$ P_\xi $ on tensors which maps $ f(x) $ to $ \pi_\xi f(x)\pi_\xi $ for
$ (x,\xi)\in\Omega M $, and write (2.2) in the shorter form
$$
HU=(P_\xi f)U,\quad \quad U|_{\partial_-\Omega M}=E.
                        \eqno{(2.4)}
$$
Because of the factor $ P_\xi $
and of the boundary condition on $ \partial_-\Omega M $,
the solution $ U $ to (2.4) satisfies
$$
U(x,\xi)\xi=U^*(x,\xi)\xi=\xi .
                        \eqno{(2.5)}
$$
Therefore the non-trivial part of the data (2.3) consists of the
restrictions $ \Phi[f](x,\xi)|_{T^\bot_{x,\xi}M} $ for
$ (x,\xi)\in\partial_+\Omega M $. This agrees with the above discussion
of the relationship between (1.6) and (1.8). The solution $ U $ is
continuous on $ \Omega M $ and $ C^\infty $-smooth on
$ \Omega M\setminus\Omega(\partial M) $ as one can easily prove using
the strict convexity of the boundary.

\bigskip

Concluding the section, let us mention one more inverse problem that is
not considered in the present article. Let $ gl(T^{\mathbb C}_xM) $ be
the space of all linear operators on $ T^{\mathbb C}_xM $. The operator
$ P_\xi $ participating in (2.4) is the orthogonal projection of the
space $ gl(T^{\mathbb C}_xM) $ onto the subspace
$$
\{f\in gl(T^{\mathbb C}_xM)\mid f\xi=f^*\xi=0\}.
$$
Let us introduce the smaller subspace
$$
\{f\in gl(T^{\mathbb C}_xM)\mid f\xi=f^*\xi=0,\
\mbox{tr}\,f=f^i_i=0\}
$$
and denote by $ Q_\xi $ the orthogonal projection onto the latter
subspace.  The corresponding inverse problem for the equation
$$
HU=(Q_\xi f)U,\quad \quad U|_{\partial_-\Omega M}=E
                        \eqno{(2.6)}
$$
is also of a great applied interest. To explain the physical meaning of
(2.6), let us return to equation (1.5) considered in the
three-dimensional case for a skew-Hermitian tensor $ f $. The
polarization vector $ \eta $ on (1.5) is a complex two-dimensional
vector subordinate to one real condition $ |\eta|=1 $. Therefore $ \eta
$ can be described by three real parameters. Two of these parameters
can be chosen to determine the shape and position of the polarization
ellipse on the plane $ \dot\gamma{}^\bot $, while the last parameter is
the phase of the electromagnetic wave. See section 6.1 of \cite{[mb]}
for a detailed discussion of the subject. Only the first two of these
parameters are measured in practice. Deleting the
wave phase from the data is mathematically equivalent to replacing the
operator $ P_\xi $ with $ Q_\xi $. The authors intend to consider the
corresponding inverse problem for (2.6) in a subsequent paper.

\section[]{Linear problem}

Let $ (M,g) $ be a CNTM. Choose two semibasic tensor fields
$ p,q\in C^\infty(\beta^1_1M;\Omega M) $ satisfying
$$
p^*(x,\xi)\xi=\xi,\quad q(x,\xi)\xi=\xi.
                        \eqno{(3.1)}
$$
For a tensor field $ f\in C^\infty(\tau^1_1M) $, consider the boundary
value problem on $ \Omega M $
$$
Hu=p(P_\xi f)q,\quad \quad u|_{\partial_-\Omega M}=0.
                        \eqno{(3.2)}
$$
The problem has a unique solution
$ u\in C(\beta^1_1M;\Omega M) $
and, in virtue of (3.1), the solution
satisfies
$$
u(x,\xi)\xi=u^*(x,\xi)\xi=0.
                        \eqno{(3.3)}
$$
In this section, we consider the inverse problem of recovering the
tensor field $ f $ from the data
$$
F[f]=u|_{\partial_+\Omega M}.
                        \eqno{(3.4)}
$$

The factors $ p $ and $ q $ on (3.2) are considered as weights. We will
assume the weights to be close to the unit weight $ E $ in the
following sense: the inequalities
$$
|p-E|<\varepsilon,\quad |q-E|<\varepsilon,\quad
|{\stackrel v{\nabla}}p|<\varepsilon,\quad
|{\stackrel v{\nabla}}q|<\varepsilon
                        \eqno{(3.5)}
$$
hold uniformly on $ \Omega M $ with the norm $ |\cdot| $ defined in
Section 2. The value of $ \varepsilon $ will be specified later.

Equation (3.2) is initially considered on $ \Omega M $. To get some
freedom in treating the equation, we extend it to the manifold
$ T^0M=\{(x,\xi)\in TM\mid \xi\neq 0\} $ of nonzero vectors. The
weights are assumed to be positively homogeneous of zero degree in $
\xi $
$$
p(x,t\xi)=p(x,\xi),\quad q(x,t\xi)=q(x,\xi)\quad
\mbox{for}\quad t>0.
$$
Then the right-hand side of (3.2) is positively homogeneous in $ \xi $
of zero degree because $ f $ is independent of $ \xi $. The solution $
u $ must be extended to $ T^0M $ as a homogeneous function of degree $
-1 $
$$
u(x,t\xi)=t^{-1}U(x,\xi)\quad \mbox{for}\quad t>0
$$
because the operator $ H $ increase the degree of homogeneity by 1.

Let us discuss smoothness properties of the solution $ u $. It can be
expresses by the explicit formula
$$
u(x,\xi)=\int\limits_{\tau_-(x,\xi)}^{0}
\Upsilon^{t,0}_\gamma\Big[p(\gamma(t),\dot\gamma(t))
P_{\dot\gamma(t)}f(\gamma(t))q(\gamma(t),\dot\gamma(t))\Big]dt,
$$
where $ \gamma=\gamma_{x,\xi} $ and $ \Upsilon^{t,0}_\gamma $ is the
parallel transport of tensors along the geodesic $ \gamma $ from the
point $ \gamma(t) $ to $ \gamma(0)=x $. The integrand is a smooth
function. Therefore smoothness properties of $ u $ are determined by
that of the integration limit $ \tau_-(x,\xi) $. The latter function is
$ C^\infty $-smooth on $ T^0M\setminus T(\partial M) $ but has
singularities on $ T^0(\partial M) $. Therefore some of integrals
considered below are improper and we have to verify their convergence.
The verification is performed in the same way as in Section 4.6 of
\cite{[mb]}. So, in order to simplify the presentation, we will pay no
attention to these singularities.

Besides (3.5), we will impose some smallness condition on the curvature
of $ (M,g) $. For $ (x,\xi)\in\Omega M $, let $ K(x,\xi) $ be the
supremum of the absolute values of sectional curvatures at the point $
x $ over all two-dimensional subspaces of $ T_xM $ containing $ \xi $.
Define
$$
k(M,g)=\sup\limits_{(x,\xi)\in\partial_-\Omega M}
\int\limits_{0}^{\tau_+(x,\xi)}
tK(\gamma_{x,\xi}(t),\dot\gamma_{x,\xi}(t))dt.
                        \eqno{(3.6)}
$$

\begin{Theorem}
For any $ n\geq 4 $, there exist positive numbers $ \delta=\delta(n) $
and $ \varepsilon=\varepsilon(n) $ such that, for any $ n $-dimensional
CNTM $ (M,g) $ satisfying
$$
k(M,g)<\delta
                        \eqno{(3.7)}
$$
and for any weights $ p,q\in C^\infty(\beta^1_1M;\Omega M) $ satisfying
(3.1) and (3.5), every tensor field $ f\in C^\infty(\tau^1_1M) $ can be
uniquely recovered from the trace (3.4) of the solution to the
boundary value problem (3.2) and the stability estimate
$$
\|f\|_{L^2}\leq C\|F[f]\|_{H^1}
                        \eqno{(3.8)}
$$
holds with a constant $ C $ independent of $ f $. In the case of
$ n=3 $, the same statement is true for a symmetric tensor field $ f $.
\end{Theorem}

In the case of a real symmetric
$ f $ and unit weights, this theorem is a
partial case of Theorem 5.2.2 of \cite{[mb]}. We will show that the
same proof works with some modifications for Theorem 3.1.

{\bf Proof of Theorem 3.1.}
We rewrite equation (3.2) in the form
$$
Hu=P_\xi f+r,
                        \eqno{(3.9)}
$$
where
$$
r=(p-E)P_\xi f+pP_\xi f(q-E).
                        \eqno{(3.10)}
$$
The remainder $ r $ is small by (3.5).
Because of (3.3), the function $ u=(u_{ij}(x,\xi)) $ is orthogonal to $
\xi $ in both indices
$$
\xi^iu_{ij}=\xi^ju_{ij}=0.
                        \eqno{(3.11)}
$$

We write down the Pestov identity for the semibasic tensor field $ u $
(see Lemma 4.4.1 of \cite{[mb]} for the case of a real
$ u $ and Lemma 5.1 of \cite{[dc]} for the general case)
$$
2\,\mbox{Re}\langle
{\stackrel h{\nabla}}u,{\stackrel v{\nabla}}Hu\rangle
=
|{\stackrel h{\nabla}}u|^2+
{\stackrel h{\nabla}}_{\!i}v^i+
{\stackrel v{\nabla}}_{\!i}w^i-
{\cal R}_1[u],
                        \eqno{(3.12)}
$$
where $ \langle\cdot,\cdot\rangle $ and $ |\cdot| $ are the
scalar product and norm on semibasic tensors defined in Section~2,
$$
v^i=\mbox{Re}\,\Big(
\xi^i{\stackrel h{\nabla}}{}^ju^{i_1i_2}\cdot
{\stackrel v{\nabla}}_{\!j}{\bar u}_{i_1i_2}-
\xi^j{\stackrel v{\nabla}}{}^i u^{i_1i_2}\cdot
{\stackrel h{\nabla}}_{\!j}{\bar u}_{i_1i_2}\Big) ,
                        \eqno{(3.13)}
$$
$$
w^i=\mbox{Re}\,\Big(
\xi^j{\stackrel h{\nabla}}{}^iu^{i_1i_2}\cdot
{\stackrel h{\nabla}}_{\!j}{\bar u}_{i_1i_2}\Big),
                        \eqno{(3.14)}
$$
$$
{\cal R}_1[u]=
R_{kplq}\xi^p\xi^q{\stackrel v{\nabla}}{}^ku^{i_1i_2}\cdot
{\stackrel v{\nabla}}{}^l{\bar u}_{i_1i_2}+
\mbox{Re}\,\Big(
(R^{i_1}_{pqj}u^{pi_2}+R^{i_2}_{pqj}u^{i_1p})
\xi^q{\stackrel v{\nabla}}{}^j{\bar u}_{i_1i_2}\Big),
                        \eqno{(3.15)}
$$
and $ (R_{ijkl}) $ is the curvature tensor.
Identity (3.12) holds on any open $ D\subset TM $ for every
$ u\in C^2(\beta^1_1M;D) $. In our case,
$ D=T^0M\setminus T(\partial M) $ and
$ u\in C^\infty(\beta^1_1M;D) $ as is shown above.

The most part of the proof deals with the left-hand side of (3.12).
We will first transform it by distinguishing some divergent terms and
then will estimate it.

From (3.9)
$$
\langle {\stackrel h{\nabla}}u,{\stackrel v{\nabla}}Hu\rangle=
\langle {\stackrel h{\nabla}}u,{\stackrel v{\nabla}}(P_\xi f)\rangle
+\langle {\stackrel h{\nabla}}u,{\stackrel v{\nabla}}r\rangle.
                        \eqno{(3.16)}
$$

We will first investigate the first term on the right-hand side of
(3.16). To this end we represent $ f $ as
$$
f_{ij}(x)={\tilde f}_{ij}(x,\xi)+
\xi_ja_i(x,\xi)+\xi_i{\bar b}_j(x,\xi)+\xi_i\xi_jc(x,\xi),
                        \eqno{(3.17)}
$$
where $ ({\tilde f}_{ij}) $ is a semibasic tensor field orthogonal to
$ \xi $ in both indices
$$
{\tilde f}_{ij}\xi^i={\tilde f}_{ij}\xi^j=0,
                        \eqno{(3.18)}
$$
semibasic covector fields $ a $ and $ b $ are orthogonal to $ \xi $
$$
a_i\xi^i=b_i\xi^i=0,
                        \eqno{(3.19)}
$$
and $ c(x,\xi) $ is a scalar function. One can easily check the
existence and uniqueness of the representation. The
(vector versions of the) fields $ a $ and
$ b $ are expressed through $ f $ by the formulas
$$
a=\frac {1} {|\xi|^2}\pi_\xi f\xi,\quad
b=\frac {1} {|\xi|^2}\pi_\xi f^*\xi.
                        \eqno{(3.20)}
$$
As follows from (3.17)--(3.19),
$$
P_\xi f=\tilde f
$$
or in coordinates
$$
(P_\xi f)_{ij}={\tilde f}_{ij}=
f_{ij}-a_i\xi_j-{\bar b}_j\xi_i-c\xi_i\xi_j.
                        \eqno{(3.21)}
$$
Differentiating the last equality with respect to $ \xi $ and using
the fact that $ f $ is independent of $ \xi $, we obtain
$$
{\stackrel v{\nabla}}_{\!k}(P_\xi f)_{ij}=
-\xi_j{\stackrel v{\nabla}}_{\!k}a_i-
\xi_i{\stackrel v{\nabla}}_{\!k}{\bar b}_j-
\xi_i\xi_j{\stackrel v{\nabla}}_{\!k}c-
g_{jk}a_i-g_{ik}{\bar b}_j-(g_{ik}\xi_j+g_{jk}\xi_i)c.
$$
Therefore
$$
\langle {\stackrel h{\nabla}}u,{\stackrel v{\nabla}}
(P_\xi f)\rangle=
{\stackrel h{\nabla}}{}^ku^{ij}\cdot
{\stackrel v{\nabla}}_{\!k}(P_\xi\bar f)_{ij}
=
$$
$$
=
{\stackrel h{\nabla}}{}^ku^{ij}
\Big(
-\xi_j{\stackrel v{\nabla}}_{\!k}{\bar a}_i-
\xi_i{\stackrel v{\nabla}}_{\!k}b_j-
\xi_i\xi_j{\stackrel v{\nabla}}_{\!k}\bar c-
g_{jk}{\bar a}_i-g_{ik}b_j-(g_{ik}\xi_j+g_{jk}\xi_i)\bar c\Big).
$$
The tensor $ {\stackrel h{\nabla}}{}^ku^{ij} $ is orthogonal to
$ \xi $ in the indices $ i $ and $ j $ as follows from (3.11).
Therefore the last formula is simplified
to the following one:
$$
\langle{\stackrel h{\nabla}}u,{\stackrel v{\nabla}}(P_\xi f)\rangle=
-{\stackrel h{\nabla}}{}^pu_{ip}\cdot\bar a{}^i
-{\stackrel h{\nabla}}{}^pu_{pi}\cdot b^i.
$$
Introducing the semibasic covector fields
$ {\stackrel h\delta}_1u $ and $ {\stackrel h\delta}_2u $ by the
equalities
$$
({\stackrel h\delta}_1u)_i={\stackrel h{\nabla}}{}^pu_{ip},\quad
({\stackrel h\delta}_2u)_i={\stackrel h{\nabla}}{}^pu_{pi},
                        \eqno{(3.22)}
$$
we write the result in the form
$$
\langle {\stackrel
h{\nabla}}u,{\stackrel v{\nabla}}(P_\xi f)\rangle= -\langle{\stackrel
h\delta}_1u,a\rangle - \langle{\stackrel h\delta}_2u,\bar b\rangle.
                        \eqno{(3.23)}
$$
This implies the estimate
$$
2\,\mbox{Re}\langle {\stackrel h{\nabla}}u,{\stackrel
v{\nabla}}(P_\xi f)\rangle \leq \frac {\beta} {2}(|{\stackrel
h\delta}_1u|^2+ |{\stackrel h\delta}_2u|^2)+ \frac {2}
{\beta}(|a|^2+|b|^2),
                      \eqno{(3.24)}
$$
where $ \beta $ is an arbitrary
positive number.

Next, we transform the expression $ |{\stackrel h\delta}_1u|^2 $ by
distinguishing a divergent term
$$
|{\stackrel h\delta}_1u|^2=
({\stackrel h\delta}_1u)^i({\stackrel h\delta}_1\bar u)_i=
{\stackrel h{\nabla}}_{\!p}u^{ip}\cdot
{\stackrel h{\nabla}}{}^q{\bar u}_{iq}
=
$$
$$
=
{\stackrel h{\nabla}}_{\!p}(u^{ip}
{\stackrel h{\nabla}}{}^q{\bar u}_{iq})-
u^{ip}{\stackrel h{\nabla}}_{\!p}
{\stackrel h{\nabla}}{}^q{\bar u}_{iq}
=
{\stackrel h{\nabla}}_{\!p}(u^{ip}
{\stackrel h{\nabla}}{}^q{\bar u}_{iq})-
u^{\cdot p}_{i\cdot}{\stackrel h{\nabla}}_{\!p}
{\stackrel h{\nabla}}_{\!q}\bar u{}^{iq}.
                        \eqno{(3.25)}
$$
By the commutator formula for horizontal derivatives (see Theorem 3.5.2
of \cite{[mb]}),
$$
{\stackrel h{\nabla}}_{\!p}
{\stackrel h{\nabla}}_{\!q}\bar u{}^{iq}=
{\stackrel h{\nabla}}_{\!q}
{\stackrel h{\nabla}}_{\!p}\bar u{}^{iq}
-R^k_{jpq}\xi^j{\stackrel v{\nabla}}_{\!k}\bar u{}^{iq} +
R^i_{jpq}\bar u{}^{jq}+R^q_{jpq}\bar u{}^{ij}.
$$
Substituting this value into the previous formula, we obtain
$$
|{\stackrel h\delta}_1u|^2=\mbox{Re}\,\Big(
-u^{\cdot p}_{i\cdot}{\stackrel h{\nabla}}_{\!q}
{\stackrel h{\nabla}}_{\!p}\bar u{}^{iq}
+{\stackrel h{\nabla}}_{\!p}(u^{ip}
{\stackrel h{\nabla}}{}^q{\bar u}_{iq})\Big)
+{\cal R}_2[u],
                        \eqno{(3.26)}
$$
where
$$
{\cal R}_2[u]=\mbox{Re}\,\Big(u^{\cdot p}_{i\cdot}
(R^k_{jpq}\xi^j{\stackrel v{\nabla}}_{\!k}\bar u{}^{iq} -
R^i_{jpq}\bar u{}^{jq}-R^q_{jpq}\bar u{}^{ij})\Big).
$$

We now transform the first summand on the right-hand side of (3.26) in
the order reverse to that used in (3.25)
$$
|{\stackrel h\delta}_1u|^2=\mbox{Re}\,\Big(
-{\stackrel h{\nabla}}_{\!q}(u^{\cdot p}_{i\cdot}
{\stackrel h{\nabla}}_{\!p}\bar u{}^{iq})+
{\stackrel h{\nabla}}_{\!q}u^{\cdot p}_{i\cdot}\cdot
{\stackrel h{\nabla}}_{\!p}\bar u{}^{iq}+
{\stackrel h{\nabla}}_{\!p}(u^{ip}
{\stackrel h{\nabla}}{}^q{\bar u}_{iq})\Big)+{\cal R}_2[u].
$$
Introducing the semibasic vector field $ {\tilde v}_1 $ by the formula
$$
({\tilde v}_1)^i=\mbox{Re}\,\Big(
u^{ji}{\stackrel h{\nabla}}{}^k{\bar u}_{jk}-
u_{jk}{\stackrel h{\nabla}}{}^k{\bar u}{}^{ji}\Big),
                        \eqno{(3.27)}
$$
we write the result in the form
$$
|{\stackrel h\delta}_1u|^2=\mbox{Re}\,\Big(
{\stackrel h{\nabla}}{}^iu^{jk}\cdot
{\stackrel h{\nabla}}_{\!k}{\bar u}_{ji}\Big)+
{\stackrel h{\nabla}}_{\!i}({\tilde v}_1)^i+{\cal R}_2[u].
                        \eqno{(3.28)}
$$

In the same way, we obtain
$$
|{\stackrel h\delta}_2u|^2=\mbox{Re}\,\Big(
{\stackrel h{\nabla}}{}^iu^{jk}\cdot
{\stackrel h{\nabla}}_{\!j}{\bar u}_{ik}\Big)+
{\stackrel h{\nabla}}_{\!i}({\tilde v}_2)^i+{\cal R}_3[u]
                        \eqno{(3.29)}
$$
with
$$
({\tilde v}_2)^i=\mbox{Re}\,\Big(
u^{ij}{\stackrel h{\nabla}}{}^k{\bar u}_{kj}-
u_{kj}{\stackrel h{\nabla}}{}^k\bar u{}^{ij}\Big)
                        \eqno{(3.30)}
$$
and
$$
{\cal R}_3[u]=\mbox{Re}\,\Big(
u^{p\cdot}_{\cdot i}(
R^k_{jpq}\xi^j{\stackrel v{\nabla}}_{\!k}\bar u{}^{qi}-
R^q_{jpq}\bar u{}^{ji}-R^i_{jpq}\bar u{}^{qi})\Big).
$$
Taking the sum of (3.28) and (3.29), we have
$$
|{\stackrel h\delta}_1u|^2+|{\stackrel h\delta}_2u|^2=
\mbox{Re}\,\Big(
{\stackrel h{\nabla}}{}^iu^{jk}\cdot
{\stackrel h{\nabla}}_{\!j}{\bar u}_{ik}+
{\stackrel h{\nabla}}{}^iu^{jk}\cdot
{\stackrel h{\nabla}}_{\!k}{\bar u}_{ji}\Big)+
{\stackrel h{\nabla}}_{\!i}{\tilde v}{}^i+{\cal R}_4[u],
                        \eqno{(3.31)}
$$
where
$$
\tilde v={\tilde v}_1+{\tilde v}_2,\quad
{\cal R}_4[u]={\cal R}_2[u]+{\cal R}_3[u].
$$

Introduce the semibasic tensor field $ z=(z_{ijk}) $ by the formula
$$
{\stackrel h{\nabla}}_{\!i}u_{jk}=
\frac {\xi_i} {|\xi|^2}(Hu)_{jk}+z_{ijk}.
                        \eqno{(3.32)}
$$
The idea of this new notation is that the tensor $ z $ is orthogonal to
$ \xi $ in all its indices
$$
\xi^iz_{ijk}=\xi^jz_{ijk}=\xi^kz_{ijk}=0,
                        \eqno{(3.33)}
$$
while the tensor
$ {\stackrel h{\nabla}}u=({\stackrel h{\nabla}}_{\!i}u_{jk}) $
has the mentioned property only in the last two indices. The summands
on the right-hand side of (3.32) are orthogonal to each other, so
$$
|{\stackrel h{\nabla}}u|^2=\frac{1}{|\xi|^2}|Hu|^2+|z|^2.
                        \eqno{(3.34)}
$$
The first two terms on the right-hand side of (3.31) can be expressed
through $ z $. Indeed, one easily see with the help of (3.32)
and (3.33) that
$$
\mbox{Re}\,\Big({\stackrel h{\nabla}}{}^iu^{jk}\cdot
{\stackrel h{\nabla}}_{\!j}{\bar u}_{ik}+
{\stackrel h{\nabla}}{}^iu^{jk}\cdot
{\stackrel h{\nabla}}_{\!k}{\bar u}_{ji}\Big)=
\mbox{Re}\,(
z^{ijk}{\bar z}_{jik}+z^{ijk}{\bar z}_{kji})\leq 2|z|^2.
$$
With the help of the last inequality, (3.31) implies the estimate
$$
|{\stackrel h\delta}_1u|^2+|{\stackrel h\delta}_2u|^2\leq
2|z|^2+
{\stackrel h{\nabla}}_{\!i}{\tilde v}{}^i+{\cal R}_4[u]
                        \eqno{(3.35)}
$$
which, together with (3.24), gives
$$
2\,\mbox{Re}
\langle{\stackrel h{\nabla}}u,{\stackrel v{\nabla}}(P_\xi f)\rangle
\leq
\beta|z|^2+\frac {2} {\beta}(|a|^2+|b|^2)+
\frac {\beta } {2}{\stackrel h{\nabla}}_{\!i}{\tilde v}{}^i
+\frac {\beta} {2}{\cal R}_4[u] .
$$
Substitute values (3.20) for $ a $ and $ b $ into the last formula
$$
2\,\mbox{Re}
\langle{\stackrel h{\nabla}}u,{\stackrel v{\nabla}}(P_\xi f)\rangle
\leq
\beta|z|^2+\frac {2} {\beta|\xi|^4}
(|\pi_\xi f\xi|^2+|\pi_\xi f^*\xi|^2)+
\frac {\beta } {2}{\stackrel h{\nabla}}_{\!i}{\tilde v}{}^i
+\frac {\beta} {2}{\cal R}_4[u] .
                        \eqno{(3.36)}
$$

Next, we estimate the second term on the right-hand side of (3.16). We
differentiate equality (3.10) with respect to $ \xi $ taking the
independence $ f $ of $ \xi $ into account
$$
{\stackrel v{\nabla}}r
= {\stackrel v{\nabla}}\left((p-E)P_\xi\right)f+ {\stackrel
v{\nabla}}(pP_\xi)f(q-E)+ pP_\xi f{\stackrel v{\nabla}}q.
$$
In what
follows in the proof, we denote different constants depending only on $
n=\mbox{dim}\,M $ by the same letter $ C $. On using (3.5), we obtain
from the last formula
$$
|{\stackrel v{\nabla}}r|\leq \frac {C\varepsilon}
{|\xi|}|f|.
$$
From this
$$
2\,\mbox{Re}\langle{\stackrel h{\nabla}}u,
{\stackrel v{\nabla}}r\rangle\leq
{C\varepsilon} (
|{\stackrel h{\nabla}}u|^2+\frac{1}{|\xi|^2}|f|^2).
                        \eqno{(3.37)}
$$

Combining (3.36) and (3.37), we obtain from (3.16)
$$
2\,\mbox{Re}\langle{\stackrel h{\nabla}}u,
{\stackrel v{\nabla}}Hu\rangle\leq
\beta|z|^2+\frac {2} {\beta|\xi|^4}
(|\pi_\xi f\xi|^2+|\pi_\xi f^*\xi|^2)+
\frac {\beta } {2}{\stackrel h{\nabla}}_{\!i}{\tilde v}{}^i
+
{C\varepsilon}(
|{\stackrel h{\nabla}}u|^2+\frac{1}{|\xi|^2}|f|^2)
+
\frac {\beta} {2}{\cal R}_4[u] .
                        \eqno{(3.38)}
$$

Estimating the left-hand side of the Pestov identity (3.12) by (3.38),
we obtain for $ |\xi|=1 $
$$
|{\stackrel h{\nabla}}u|^2
+
{\stackrel v{\nabla}}_{\!i}w^i-\beta|z|^2
-
\frac {2} {\beta}(|\pi_\xi f\xi|^2+|\pi_\xi f^*\xi|^2)
-
C\varepsilon(|{\stackrel h{\nabla}}u|^2+|f|^2)
\leq
{\stackrel h{\nabla}}_{\!i}(\frac {\beta} {2}\tilde v{}^i-v^i)
+{\cal R}[u],
                        \eqno{(3.39)}
$$
where
$$
{\cal R}[u]={\cal R}_1[u]+\frac {\beta} {2}{\cal R}_4[u].
$$
We multiply inequality (3.39) by the volume form
$ d\Sigma=|d_x\omega(\xi)\wedge dV^n(x)| $, integrate over
$ \Omega M $, and transform the integrals of divergent terms by
Gauss-Ostrogradskii formulas (see Theorem 3.6.3 of \cite{[mb]})
$$
\int\limits_{\Omega M}
\Big[
|{\stackrel h{\nabla}}u|^2
+
(n-2)|Hu|^2-\beta|z|^2
-
\frac {2} {\beta}(|\pi_\xi f\xi|^2+|\pi_\xi f^*\xi|^2)
-
C\varepsilon(|{\stackrel h{\nabla}}u|^2+|f|^2)
\Big]d\Sigma
\leq
$$
$$
\leq
\int\limits_{\partial\Omega M} \langle \frac {\beta} {2}\tilde
v-v,\nu\rangle d\Sigma^{2n-2} +\int\limits_{\Omega M}{\cal
R}[u]d\Sigma .
                        \eqno{(3.40)}
$$
The second term on the left-hand side has appeared because $ w $ is
positively homogeneous of degree $ -1 $ in $ \xi $ and satisfies
$ \langle \xi,w\rangle=|Hu|^2 $ as is seen from (3.14). Substituting
the value $ |{\stackrel h{\nabla}}u|^2=|z|^2+|Hu|^2 $ from (3.34), we
write the result in the form
$$
\int\limits_{\Omega M}
\Big[
(1-\beta-C\varepsilon)|z|^2
+
(n-1-C\varepsilon)|Hu|^2
-
\frac {2} {\beta}(|\pi_\xi f\xi|^2+|\pi_\xi f^*\xi|^2)
-
C\varepsilon |f|^2
\Big]d\Sigma
\leq
$$
$$
\leq
\int\limits_{\partial\Omega M} \langle \frac {\beta} {2}\tilde
v-v,\nu\rangle d\Sigma^{2n-2} +\int\limits_{\Omega M}{\cal
R}[u]d\Sigma .
                        \eqno{(3.41)}
$$

Assuming $ \beta\leq 1 $, integrals on the right-hand side of (3.41)
can be estimated exactly as in Section 5.5 of \cite{[mb]}
$$
\left|\int\limits_{\Omega M}{\cal R}[u]d\Sigma \right|\leq
Ck(M,g)\int\limits_{\Omega M}
|{\stackrel h{\nabla}}u|^2d\Sigma
\leq
C\delta \int\limits_{\Omega M}
|{\stackrel h{\nabla}}u|^2d\Sigma,
                        \eqno{(3.42)}
$$
$$
\left|
\int\limits_{\partial\Omega M} \langle \frac {\beta} {2}\tilde
v-v,\nu\rangle d\Sigma^{2n-2}
\right|\leq
D\|u|_{\partial_+\Omega M}\|_{H^1}^2,
                        \eqno{(3.43)}
$$
where $ k(M,g) $ is defined by (3.6). The second inequality on (3.42)
is valid because of (3.7). The constant $ D $ on (3.43) depends on
$ (M,g) $, unlike the constant $ C $ on (3.42) which depends only on $
n $.

Combining (3.41)--(3.43) and using again the equality
$ |{\stackrel h{\nabla}}u|^2=|z|^2+|Hu|^2 $, we obtain
$$
\int\limits_{\Omega M}
\Big[
(1-\beta-C\varepsilon-C\delta)|z|^2
+
(n-1-C\varepsilon-C\delta)|Hu|^2
-
\frac {2} {\beta}(|\pi_\xi f\xi|^2+|\pi_\xi f^*\xi|^2)
-
C\varepsilon |f|^2
\Big]d\Sigma
\leq
$$
$$
\leq D\|u|_{\partial_+\Omega M}\|_{H^1}^2
                        \eqno{(3.44)}
$$
with some new constant $ C $ depending only on $ n $.

Let us compare $ |Hu| $ and $ |P_\xi f| $. The estimate
$ |r|<C\varepsilon|f| $ follows from (3.5) and (3.10).
The latter, together with (3.9), implies
$$
|Hu|^2\geq|P_\xi f|^2-C\varepsilon|f|^2.
                        \eqno{(3.45)}
$$
Using the last inequality, we transform (3.44) to the final form
$$
\int\limits_{M}\int\limits_{\Omega_xM}
\Big[
(1-\beta-C\varepsilon-C\delta)|z|^2
+
(n-1-C\varepsilon-C\delta)|P_\xi f|^2
- \quad \quad \quad \quad \quad \quad
$$
$$
-
\frac {2} {\beta}(|\pi_\xi f\xi|^2+|\pi_\xi f^*\xi|^2)
-
C\varepsilon |f|^2
\Big]d\omega_x(\xi)dV^n(x)
\leq
D\|u|_{\partial_+\Omega M}\|_{H^1}^2.
                        \eqno{(3.46)}
$$
Let us remind that $ \beta $ is an arbitrary number satisfying
$ 0<\beta\leq 1 $.

\begin{Lemma}
For every Riemannian manifold $ (M,g) $ of dimension $ n\geq 4 $ and
every point $ x\in M $, the Hermitian form
$$
B(f,f)=
\int\limits_{\Omega_xM}
\Big[
(n-1)|P_\xi f|^2
-
2(|\pi_\xi f\xi|^2+|\pi_\xi f^*\xi|^2)
\Big]d\omega_x(\xi)
$$
is positive definite on the space of  second
rank tensors at $ x $. Moreover, the estimate
$$
B(f,f)\geq c|f|^2
$$
holds with a positive constant $ c $ depending only on $ n $. In the
case of $ n=3 $, the same statement is valid for symmetric
tensors.
\end{Lemma}

The proof of the lemma will be given later, and now we finish the proof
of Theorem 3.1 by making use of the lemma.

By the lemma, the inequality
$$
c\|f\|^2_{L^2}=c\int\limits_{M}|f|^2\,dV^n(x)
\leq
\int\limits_{M}\int\limits_{\Omega_xM}
\Big[
(1-\beta-C\varepsilon-C\delta)|z|^2
+
$$
$$
+
(n-1-C\varepsilon-C\delta)|P_\xi f|^2
-
\frac {2} {\beta}(|\pi_\xi f\xi|^2+|\pi_\xi f^*\xi|^2)
-
C\varepsilon |f|^2
\Big]d\omega_x(\xi)dV^n(x)
                        \eqno{(3.47)}
$$
holds for $ \beta=1,\ \varepsilon=\delta=0 $. By continuity
and by estimates $ |P_\xi f|\leq|f| $, $ |\pi_\xi f\xi|\leq|f| $,
$ |\pi_\xi f^*\xi|\leq|f| $ for $ |\xi|=1 $,
the same
inequality holds for some positive $ c,\varepsilon,\delta, $ and
$ \beta $ independent of $ f $ and satisfying
$ 1-\beta-C\varepsilon-C\delta\geq 0 $. Combining (3.46) and (3.47), we
obtain
$$
\|f\|^2_{L^2}\leq C\|u|_{\partial_+\Omega M}\|_{H^1}^2
=C\|F[f]\|_{H^1}^2
$$
with $ C=D/c $. This finishes the proof of
Theorem 3.1.

\bigskip

{\bf Proof of Lemma 3.2.}
One easily checks the equality
$ B(f,f)=B(u,u)+B(v,v) $ for a complex tensor $ f $ represented as
$ f=u+iv $ with real $ u $ and $ v $. Therefore it suffices to prove
the statement for a real tensor $ f $.

The corresponding symmetric bilinear form is
$$
B(f,h)=\int\limits_{\Omega_xM}\Big[
(n-1)\langle P_\xi f,h\rangle -
2(\langle\pi_\xi f\xi,h\xi\rangle
+\langle\pi_\xi f^*\xi,h^*\xi\rangle)\Big]d\omega_x(\xi).
$$
Obviously, $ B(f,h)=0 $ for a symmetric $ f $ and skew-symmetric $ h $.
Therefore, it suffices to prove the positiveness of $ B $ on the spaces
of real symmetric and skew-symmetric tensors separately.

The positiveness of $ B $ on the space of real symmetric tensors is
proved in Lemma 5.6.1 of \cite{[mb]}, where $ \pi_\xi f\xi $ is denoted
by $ P_\xi j_\xi f $.

Thus, we have to consider the quadratic form $ B $ on the space of real
skew-symmetric tensors. On making use of an orthonormal basis,
we identify $ T_xM $ with $ {\mathbb R}^n $ endowed with the
standard scalar product and identify $ \Omega_xM $ with the unit sphere
$ \Omega\subset{\mathbb R}^n $. So
$$
\frac {1} {\omega}B(f,f)=
\frac {1} {\omega}\int\limits_{\Omega}
\Big[(n-1)|P_\xi f|^2-4|\pi_\xi f\xi|^2\Big]d\omega(\xi),
                        \eqno{(3.48)}
$$
where $ \omega $ is the volume of $ \Omega $ and $ d\omega $ is the
standard volume form on $ \Omega  $.

For a skew-symmetric $ f $
$$
\pi_\xi f\xi =f\xi
                        \eqno{(3.49)}
$$
since $ f\xi $ is orthogonal to $ \xi $. Therefore formulas (3.20) and
(3.21) are simplified to the following ones:
$$
a=-b=f\xi,\quad (P_\xi f)_{ij}=f_{ij}-(f\xi)_i\xi_j+
\xi_i(f\xi)_j
$$
for $ |\xi|=1 $. Using the last equality and
$ \langle f\xi,\xi\rangle=0 $, we easily calculate
$$
|P_\xi f|^2=|f|^2-2|f\xi|^2\quad \mbox{for}\quad |\xi|=1.
                        \eqno{(3.50)}
$$
Substitute (3.49) and (3.50) into (3.48)
$$
\frac {1} {\omega}B(f,f)=
(n-1)|f|^2-
\frac {2(n+1)} {\omega}\int\limits_{\Omega}
|f\xi|^2\,d\omega.
                        \eqno{(3.51)}
$$
On using the obvious relation
$$
\frac {1} {\omega}\int\limits_{\Omega}\xi_i\xi_j\,d\omega=
\left\{
\begin{array}{l}
0\quad \mbox{for}\quad i\neq j,\\
1/n\quad \mbox{for}\quad i=j,
\end{array}\right.
$$
we find
$$
\frac {1} {\omega}\int\limits_{\Omega}
|f\xi|^2\,d\omega=\frac {1} {n}|f|^2.
$$
Inserting this value into (3.51), we see that
$$
\frac {1} {\omega}B(f,f)=
\frac {n^2-3n-2} {n}|f|^2.
$$
This implies the positiveness of $ B $ on skew-symmetric tensors for
$ n\geq 4 $.

\section[]{Three-dimensional case}

We will first show that both our problems, linear and nonlinear,
possess some non-uniqueness in the three-dimensional case.

Let $ (M,g) $ be a three-dimensional CNTM which is assumed to be
oriented. Every tangent space $ T_xM $ is a three-dimensional oriented
Euclidean space. So, the vector product
$$
T_xM\times T_xM\rightarrow T_xM,\quad
(v,w)\mapsto v\times w
$$
is well defined. It is extended to the $ \mathbb C $-bilinear operation
$$
T^{\mathbb C}_xM\times T^{\mathbb C}_xM
\rightarrow T^{\mathbb C}_xM,\quad
(v,w)\mapsto v\times w
$$
on complex vectors. For a complex vector field
$ v\in C^\infty(\tau^1_0M) $, we denote by
$ L_v\in C^\infty(\tau^1_1M) $ the operator of vector multiplication by
$ v $,
$$
L_v(x)\eta=v(x)\times\eta\quad\mbox{for}\quad\eta\in T^{\mathbb C}_xM.
$$
Note that $ L_v $ is a skew-symmetric tensor field. Quite similarly,
for a semibasic vector field $ v\in C^\infty(\beta^1_0M) $, the
operator $ L_v\in C^\infty(\beta^1_1M) $ is defined. Let us prove the
formula
$$
P_\xi L_v=\pi_\xi L_v\pi_\xi=
\frac {\langle v,\xi\rangle} {|\xi|^2}L_\xi.
                        \eqno{(4.1)}
$$
We remind that $ \langle\cdot,\cdot\rangle $ and $ |\cdot| $ are
defined in Section 2.
This is a pure algebraic local formula. So, we can use a positive
orthonormal basis $ (e_1,e_2,e_3=\xi/|\xi|) $ in $ T_xM $. In such a
basis
$$
\pi_\xi=
\left(
\begin{array}{ccc}
1&0&0\\
0&1&0\\
0&0&0
\end{array}\right),\
L_v=
\left(
\begin{array}{ccc}
0&-v_3&v_2\\
v_3&0&-v_1\\
-v_2&v_1&0
\end{array}\right),\
L_\xi=
\left(
\begin{array}{ccc}
0&-|\xi|&0\\
|\xi|&0&0\\
0&0&0
\end{array}\right),\
\langle v,\xi\rangle=v_3|\xi|,
$$
and the formula follows immediately.

Next, we prove the formula
$$
{\stackrel h{\nabla}}L_\xi =0.
                        \eqno{(4.2)}
$$
The formula is quite expectable since
$ {\stackrel h{\nabla}}\xi =0 $. Nevertheless, it needs a proof. Here,
we have to use a general coordinate system since $ \stackrel h{\nabla}
$ is a differential operator. The vector product is expressed by the
formula
$$
(v\times w)^i=\frac {1} {\sqrt{g}}
(v_{i+1}w_{i+2}-v_{i+2}w_{i+1})
$$
in general coordinates, where
$ g=\mbox{det}(g_{ij}) $ and indices are reduced modulo 3. Therefore
$$
(L_v)_{i,i+1}=-\sqrt{g}v^{i+2}
                        \eqno{(4.3)}
$$
and (4.2) is proved by straightforward calculations with making use of
$ {\nabla}_{\!i}g_{jk}=0 $ and
$ {\stackrel h{\nabla}}_{\!i}\xi^j=0 $.

We derive from (4.1)--(4.2) and the formula $ H(1/|\xi|^2)=0 $ that
$$
H(\frac {\lambda} {|\xi|^2}L_\xi) =
\frac {H\lambda} {|\xi|^2}L_\xi=P_\xi L_{{\nabla}\lambda}
$$
for a complex function $ \lambda\in C^\infty(M) $.
Thus
$$
H(\frac {\lambda} {|\xi|^2}L_\xi) =
P_\xi L_{{\nabla}\lambda}.
                            \eqno{(4.4)}
$$
This gives us the following non-uniqueness in the linear problem. If
the function $ \lambda $ vanishes on the boundary,
$ \lambda|_{\partial M}=0 $, then
$ u(x,\xi)=\lambda L_\xi/|\xi|^2 $ solves the boundary value problem
$$
Hu=P_\xi f,\quad u_{\partial_-\Omega M}=0
                        \eqno{(4.5)}
$$
with $ f=L_{{\nabla}\lambda} $ and satisfies
$$
F[f]=u|_{\partial_+\Omega M}=0.
                        \eqno{(4.6)}
$$
The boundary value problem (4.5) coincides with (3.2) for the unit
weights $ p=q=E $, and (4.6) means that $ f $ cannot be recovered from
$ F[f] $.

The same arguments give us some non-uniqueness in the nonlinear
problem. Fix a function $ \lambda\in C^\infty(M) $ vanishing on the
boundary, $ \lambda|_{\partial M}=0 $, and, for
$ (x,\xi)\in\Omega M $, define the linear operator $ U(x,\xi) $ on
$ T_x^{\mathbb C}M $ whose matrix in a positive orthonormal basis
$ (e_1,e_2,e_3=\xi) $ is
$$
\left(
\begin{array}{ccc}
\cos\lambda(x)&-\sin\lambda(x)&0\\
\sin\lambda(x)&\cos\lambda(x)&0\\
0&0&1
\end{array}\right).
$$
For a real function $ \lambda $, $ U(x,\xi) $ is the rotation of $ T_xM
$ around the axis $ \xi $ by the angle $ \lambda(x) $. In the case of a
complex $ \lambda $, the operator is also well defined although its
geometric sense is more complicated. The semibasic tensor field
$ U\in C^\infty(\beta^1_1M;\Omega M) $ satisfies the equation
$$
HU=(P_\xi L_{{\nabla}\lambda})U
                        \eqno{(4.7)}
$$
and boundary condition
$$
U|_{\partial\Omega M}=E.
                        \eqno{(4.8)}
$$
Indeed, (4.8) is obvious. Let us prove (4.7). In virtue of (4.1),
equation (4.7) is equivalent to the following one:
$$
HU(x,\xi)=
\langle{\nabla}\lambda(x),\xi\rangle L_\xi U(x,\xi).
                        \eqno{(4.9)}
$$
Let $ \gamma $ be a unit speed geodesic. Setting
$ x=\gamma(t) $, $ \xi=\dot\gamma(t) $ in (4.9), we arrive to the
equation
$$
\frac {DU(t)} {dt}=\frac {d\lambda(t)} {dt}
L_{\dot\gamma(t)}U(t),
                        \eqno{(4.10)}
$$
where $ \lambda(t)=\lambda(\gamma(t)) $ and
$ U(t)=U(\gamma(t),\dot\gamma(t)) $. Conversely, if (4.10) holds for
any unit speed geodesic $ \gamma $, then (4.9) is true. To prove (4.10), we choose
an orthonormal basis $ (e_1(t),e_2(t),e_3(t)=\dot\gamma(t)) $ of
$ T_{\gamma(t)}M $ which is parallel along $ \gamma $. In such a basis,
(4.10) is equivalent to the matrix equation
$$
\frac {d} {dt}
\left(
\begin{array}{ccc}
\cos\lambda(t)&-\sin\lambda(t)&0\\
\sin\lambda(t)&\cos\lambda(t)&0\\
0&0&1
\end{array}\right)
=
\frac {d\lambda} {dt}
\left(
\begin{array}{ccc}
0&-1&0\\
1&0&0\\
0&0&0
\end{array}\right)
\left(
\begin{array}{ccc}
\cos\lambda(t)&-\sin\lambda(t)&0\\
\sin\lambda(t)&\cos\lambda(t)&0\\
0&0&1
\end{array}\right)
$$
which is obviously true.

Comparing (4.7)--(4.8) with (2.3)--(2.4), we see that a field
$ f=L_{{\nabla}\lambda} $ with $ \lambda|_{\partial M}=0 $ cannot be
recovered from data (2.3).

Let us introduce the definition: $ f\in C^\infty(\tau^1_1M) $ is said
to be a {\it potential field} if it can be represented as
$ f=L_{{\nabla}\lambda} $ for some function $ \lambda\in C^\infty(M) $
vanishing on the boundary, $ \lambda|_{\partial M}=0 $. Potential
fields constitute the natural obstruction for the uniqueness in the
both, linear and nonlinear, problems. We are going to prove that a
solution to the linear problem is unique up to the obstruction. The
corresponding local result for the nonlinear problem will be obtained
in the next section. First of all we prove

\begin{Lemma} (on decomposition).
Let $ (M,g) $ be a compact oriented three-dimensional Riemannian
manifold with boundary. Every tensor field $ f\in C^\infty(\tau^1_1M) $
can be uniquely represented as
$$
f=L_{{\nabla}\lambda}+\tilde f
                        \eqno{(4.11)}
$$
with some $ \lambda\in C^\infty(M) $ satisfying
$$
\lambda|_{\partial M}=0
                        \eqno{(4.12)}
$$
and some tensor field $ \tilde f\in C^\infty(\tau^1_1M) $ satisfying
the condition: the 2-form $ {\tilde f}_{ij}dx^i\wedge dx^j $ is closed,
$$
d({\tilde f}_{ij}dx^i\wedge dx^j)=0.
                            \eqno{(4.13)}
$$
\end{Lemma}

The summands of decomposition (4.11) are called the {\it potential} and
{\it closed} parts of $ f $ respectively. Note that (4.13) involves
only the skew-symmetric part of $ \tilde f $, i.e., a symmetric tensor
field is closed. Lemma 4.1 can be derived from the Hodge-Morrey
decomposition \cite{[S]} but we give a shorter independent proof.

{\bf Proof of Lemma 4.1.}
We first prove the uniqueness statement. Assume (4.11)--(4.13) to be
valid. Applying the exterior derivative $ d $ to the form
$ f_{ij}dx^i\wedge dx^j $ and using (4.11) and (4.13), we obtain
$$
d\Big((L_{{\nabla}\lambda})_{ij}dx^i\wedge dx^j\Big)
=d(f_{ij}dx^i\wedge dx^j).
$$
On using (4.3), one can check by a straightforward calculation in
coordinates that
$$
d\Big((L_{{\nabla}\lambda})_{ij}dx^i\wedge dx^j\Big)
=
-2{\Delta}\lambda \sqrt{g}dx^1\wedge dx^2\wedge dx^3,
$$
where $ \Delta  $ is the Laplace-Beltrami operator. Thus, the function
$ \lambda $ solves the Dirichlet problem
$$
\Delta\lambda=
-\frac {1} {\sqrt{g}}
\Big(\frac {\partial f^-_{23}} {\partial x^1}+
\frac {\partial f^-_{31}} {\partial x^2}+
\frac {\partial f^-_{12}} {\partial x^3}\Big),
\quad \lambda|_{\partial M}=0,
                        \eqno{(4.14)}
$$
where $ f^- $ is the skew-symmetric part of $ f $, i.e.,
$ f^-_{ij}=\frac {1} {2}(f_{ij}-f_{ji}) $. The solution to the
Dirichlet problem is unique.

The existence statement is proved by reverse arguments. Given $ f $,
let $ \lambda $ be the solution to the Dirichlet problem (4.14) and
$ \tilde f=f-L_{{\nabla}\lambda} $. Then (4.11)--(4.13) holds. The lemma
is proved.

We restrict ourselves to considering the inverse problem for closed
fields only.

\begin{Theorem}
There exist such positive numbers $ \delta $ and $ \varepsilon $ that,
for any oriented 3-di\-men\-si\-onal CNTM
$ (M,g) $ satisfying (3.7) and for
any weights $ p,q\in C^\infty(\beta^1_1M;\Omega M) $ satisfying (3.1)
and (3.5), every closed tensor field $ f\in C^\infty(\tau^1_1M) $ can be
uniquely recovered from the trace (3.4) of the solution to the boundary
value problem (3.2) and the stability estimate
$$
\|f\|^2_{L^2}\leq C\left(
\|F[f]\|^2_{H^1}+\|f|_{\partial M}\|_{L^2}\cdot
\|F[f]\|_{L^2}\right)
                        \eqno{(4.15)}
$$
holds with a constant $ C $ independent of $ f $.
\end{Theorem}

{\bf Proof} follows the same line as the proof of Theorem 3.1 with the
following difference: the left-hand side of the Pestov identity (3.12)
will be estimated in a different way by making use of the closeness of
$ f $.

We represent $ f $ as the sum of symmetric and skew-symmetric fields
$$
f=f^++f^-,\quad f^+_{ij}=f^+_{ji},\quad
f^-_{ij}=-f^-_{ji}.
$$
Taking the symmetry of Christoffel symbols into account, the
closeness condition for $ f $ can be written as
$$
{\nabla}_{\!i}f^-_{jk}+
{\nabla}_{\!j}f^-_{ki}+
{\nabla}_{\!k}f^-_{ij}=0.
                        \eqno{(4.16)}
$$
The vector $ f^-\xi $ is orthogonal to $ \xi $ and therefore
$ \pi_\xi f^-\xi=f^-\xi $. Formulas (3.20) take now the form
$$
a=\frac {1} {|\xi|^2}\pi_\xi f^+\xi+
\frac {1} {|\xi|^2}f^-\xi,\quad
b=\frac {1} {|\xi|^2}\pi_\xi\bar f{}^+\xi-
\frac {1} {|\xi|^2}\bar f{}^-\xi.
                        \eqno{(4.17)}
$$

We write equation (3.2) in the form (3.9) with the remainder $ r $
defined by (3.10). Then
we write the Pestov identity (3.12) for $ u $ with terms defined by
(3.13)--(3.15). The left hand-side of the identity can be written as in
(3.16).

The main problem is estimating the first term
$\langle{\stackrel h{\nabla}}u,{\stackrel v{\nabla}}(P_\xi f)\rangle $
on
the right-hand side of (3.16). To this end we represent $ f $ in
form (3.17) with $ a $ and $ b $ defined by (4.17). Then (3.23) holds.
In view of (4.17), equation (3.23) can be written as
$$
\langle{\stackrel h{\nabla}}u,{\stackrel v{\nabla}}(P_\xi f)\rangle
=-\frac {1} {|\xi|^2}\langle
{\stackrel h\delta}_1u+{\stackrel h\delta}_2u,
\pi_\xi f^+\xi\rangle -
\frac {1} {|\xi|^2}\langle
{\stackrel h\delta}_1u-{\stackrel h\delta}_2u,
f^-\xi\rangle.
                    \eqno{(4.18)}
$$

We transform the second term on the right-hand side of (4.18) by
distinguishing divergent terms. Using definition (3.22) of
$ {\stackrel h\delta}_1u $ and $ {\stackrel h\delta}_2u $, we write
$$
\langle
{\stackrel h\delta}_1u-{\stackrel h\delta}_2u,
f^-\xi\rangle
=
({\stackrel h{\nabla}}_{\!p}u^{ip}-
{\stackrel h{\nabla}}_{\!p}u^{pi})
\bar f{}^-_{ik}\xi^k
=
$$
$$
=
{\stackrel h{\nabla}}_{\!p}(
\xi^ku^{ip}\bar f{}^-_{ik}+\xi^ku^{pi}\bar f{}^-_{ki})
-
\xi^ku^{ip}({\stackrel h{\nabla}}_{\!p}\bar f{}^-_{ik}+
{\stackrel h{\nabla}}_{\!i}\bar f{}^-_{kp}).
$$
Now, we use the closeness condition (4.16) to transform this formula to
the following one:
$$
\langle
{\stackrel h\delta}_1u-{\stackrel h\delta}_2u,
f^-\xi\rangle
=
\xi^ku^{ip}{\stackrel h{\nabla}}_{\!k}\bar f{}^-_{pi}
+
{\stackrel h{\nabla}}_{\!p}(
\xi^ku^{ip}\bar f{}^-_{ik}+\xi^ku^{pi}\bar f{}^-_{ki}).
$$
Finally, we distinguish a divergent term from the first summand on the
right-hand side
$$
\langle
{\stackrel h\delta}_1u-{\stackrel h\delta}_2u,
f^-\xi\rangle
=
{\stackrel h{\nabla}}_{\!k}(\xi^ku^{ip}\bar f{}^-_{pi})
-
\xi^k{\stackrel h{\nabla}}_{\!k}u^{ip}\cdot\bar f{}^-_{pi}
+
{\stackrel h{\nabla}}_{\!p}(
\xi^ku^{ip}\bar f{}^-_{ik}+\xi^ku^{pi}\bar f{}^-_{ki}).
$$
This can be written as
$$
\langle
{\stackrel h\delta}_1u-{\stackrel h\delta}_2u,
f^-\xi\rangle
=
\langle Hu,f^-\rangle
+
{\stackrel h{\nabla}}_{\!i}(\xi^iu^{jk}\bar f{}^-_{kj}
+
\xi^ku^{ji}\bar f{}^-_{jk}
+
\xi^ku^{ij}\bar f{}^-_{kj}).
                        \eqno{(4.19)}
$$

Next, we calculate the first term on the right-hand side of (4.19) by
making use of (3.9) and (3.21)
$$
\langle Hu,f^-\rangle
=
(Hu)^{ij}\bar f{}^-_{ij}
=
((P_\xi f)^{ij}+r^{ij})\bar f{}^-_{ij}
=
(f^{ij}-a^i\xi^j-{\bar b}{}^j\xi^i-
c\xi^i\xi^j+r^{ij})\bar f{}^-_{ij}.
                        \eqno{(4.20)}
$$
Since $ f^- $ is a skew-symmetric tensor,
$ c\xi^i\xi^j\bar f{}^-_{ij}=0 $. This means that the term
$ c\xi^i\xi^j $ on the right-hand side of (4.20) can be omitted. By the
same reason, the term $ f^{ij} $ can be replaced with
$ (f^-)^{ij} $, the vector
$ a=\frac {1} {|\xi|^2}\pi_\xi f\xi $ can be replaced with
$ \frac {1} {|\xi|^2}f\xi $, and the vector $ b $ can be replaced with
$ \frac {1} {|\xi|^2}f^*\xi $. In such the way, (4.20) takes the form
$$
\langle Hu,f^-\rangle
=
|f^-|^2
-\frac {1} {|\xi|^2}
(f^{ik}\xi_k\xi^j+f^{kj}\xi_k\xi^i)\bar f{}^-_{ij}
+
\langle r,f^-\rangle .
$$
The second term on the right-hand side is independent of
$ f^+ $, and the formula can be written as
$$
\langle Hu,f^-\rangle
=
|f^-|^2
-\frac {2} {|\xi|^2}
|f^-\xi|^2+\langle r,f^-\rangle .
                        \eqno{(4.21)}
$$

Substitute (4.21) into (4.19)
$$
\langle
{\stackrel h\delta}_1u-{\stackrel h\delta}_2u,
f^-\xi\rangle
=
|f^-|^2
-\frac {2} {|\xi|^2}
|f^-\xi|^2+\langle r,f^-\rangle
+
{\stackrel h{\nabla}}_{\!i}(\xi^iu^{jk}\bar f{}^-_{kj}
+
\xi^ku^{ji}\bar f{}^-_{jk}
+
\xi^ku^{ij}\bar f{}^-_{kj}).
$$
Inserting this expression into (4.18) and estimating the first term on
the right-hand side of (4.18) in a similar way as in deriving (3.24),
we arrive to the inequality
$$
2\,\mbox{Re}
\langle{\stackrel h{\nabla}}u,{\stackrel v{\nabla}}(P_\xi f)\rangle
\leq
\frac {\beta} {2}(|{\stackrel h\delta}_1u|^2+
|{\stackrel h\delta}_2u|^2)
+
\frac {4} {\beta|\xi|^4}|\pi_\xi f^+\xi|^2
-
$$
$$
-
\frac {2} {|\xi|^2}|\bar f{}^-|^2
+\frac {4} {|\xi|^4}|f^-\xi|^2
+
{\stackrel h{\nabla}}_{\!i}({\tilde v}_3)^i
+\frac {2} {|\xi|^2}|r|\cdot|f|,
                        \eqno{(4.22)}
$$
where
$$
({\tilde v}_3)^i
=
-\frac {2} {|\xi|^2}\,\mbox{Re}(
\xi^iu^{jk}\bar f{}^-_{kj}
+
\xi^ku^{ji}\bar f{}^-_{jk}
+
\xi^ku^{ij}\bar f{}^-_{kj})
                        \eqno{(4.23)}
$$
and $ \beta $ is an arbitrary positive number.
The last term on the right-hand side of (4.22) can be estimated by
$ C\varepsilon|f|^2/|\xi|^2 $ as follows from (3.5) and (3.10).
Estimating the first term on the right-hand side of (4.22) by (3.35),
we obtain
$$
2\,\mbox{Re}
\langle{\stackrel h{\nabla}}u,{\stackrel v{\nabla}}(P_\xi f)\rangle
\leq
\beta|z|^2
+
\frac {4} {\beta|\xi|^4}|\pi_\xi f^+\xi|^2
-
\frac {2} {|\xi|^2}|\bar f{}^-|^2
+\frac {4} {|\xi|^4}|f^-\xi|^2
+
\frac {C\varepsilon} {|\xi|^2}|f|^2+
{\stackrel h{\nabla}}_{\!i}\tilde v{}^i
+\frac {\beta} {2}{\cal R}_4[u]
                        \eqno{(4.24)}
$$
with the same curvature dependent term $ {\cal R}_4[u] $ as in (3.36)
and
$$
\tilde v=\frac {\beta} {2}({\tilde v}_1+{\tilde v}_2)+
{\tilde v}_3,
$$
where $ {\tilde v}_i\ (i=1,2,3) $ are defined by (3.27), (3.30), and
(4.23) respectively.

The second term on the right-hand side of (3.16) is estimated by (3.37)
as before. Combining (3.37) and (4.24), we obtain from (3.16)
$$
2\,\mbox{Re}
\langle{\stackrel h{\nabla}}u,{\stackrel v{\nabla}}Hu\rangle
\leq
\beta|z|^2
+
\frac {4} {\beta}|\pi_\xi f^+\xi|^2
-
2|\bar f{}^-|^2
+4|f^-\xi|^2
+
C\varepsilon(|f|^2+|{\stackrel h{\nabla}}u|^2)+
{\stackrel h{\nabla}}_{\!i}\tilde v{}^i
+\frac {\beta} {2}{\cal R}_4[u]
                        \eqno{(4.25)}
$$
for $ |\xi|=1 $.

We estimate the left-hand side of the Pestov identity (3.12) by (4.25)
and write the result in the form
$$
(1-C\varepsilon) |{\stackrel h{\nabla}}u|^2
+
{\stackrel v{\nabla}}_{\!i}w^i
-
\beta|z|^2
-
\frac {4} {\beta}|\pi_\xi f^+\xi|^2
+
2|\bar f{}^-|^2
-4|f^-\xi|^2
-
C\varepsilon|f|^2
\leq
{\stackrel h{\nabla}}_{\!i}(\tilde v{}^i-v^i)
+\frac {\beta} {2}{\cal R}_4[u].
$$
Integrating this inequality and transforming the integrals of divergent
terms in the same way as in (3.40), we obtain
$$
\int\limits_{\Omega M}\Big[
(1-C\varepsilon) |{\stackrel h{\nabla}}u|^2
+
|Hu|^2-\beta|z|^2
-
\frac {4} {\beta}|\pi_\xi f^+\xi|^2
+
2|\bar f{}^-|^2
-4|f^-\xi|^2
-
C\varepsilon|f|^2
\Big]d\Sigma
\leq
$$
$$
\leq
\int\limits_{\partial\Omega M}
\langle \tilde v-v,\nu\rangle d\Sigma^{2n-2}
+
\int\limits_{\Omega M}
{\cal R}_4[u]d\Sigma .
$$
Substituting the value
$ |{\stackrel h{\nabla}}u|^2=|z|^2+|Hu|^2 $ from (3.34), we write the
result in the form
$$
\int\limits_{\Omega M}\Big[
(1-\beta-C\varepsilon)|z|^2
+
(2-C\varepsilon)|Hu|^2
-
\frac {4}{\beta}|\pi_\xi f^+\xi|^2
+
2|\bar f{}^-|^2 -4|f^-\xi|^2
-
C\varepsilon|f|^2 \Big]d\Sigma
\leq
$$
$$
\leq
\int\limits_{\partial\Omega M}
\langle \tilde v-v,\nu\rangle d\Sigma^{2n-2}
+
\int\limits_{\Omega M}
{\cal R}_4[u]d\Sigma .
                        \eqno{(4.26)}
$$

The curvature dependent integral on (4.26) is estimated as before
in (3.42)
$$
\left|\int\limits_{\Omega M}
{\cal R}_1[u]d\Sigma\right|\leq
C\delta \int\limits_{\Omega M}
|{\stackrel h{\nabla}}u|^2d\Sigma ,
                        \eqno{(4.27)}
$$
while the boundary integral is estimated in a little bit different way.
Namely, instead of (3.43), we have the estimate
$$
\left|\int\limits_{\partial\Omega M}
\langle \tilde v-v,\nu\rangle d\Sigma^{2n-2}\right|
\leq
D\Big(
\|u|_{\partial_+\Omega M}\|^2_{H^1}+
\|f|_{\partial M}\|_{L^2}\cdot
\|u|_{\partial_+\Omega M}\|_{L^2}\Big).
                        \eqno{(4.28)}
$$
The second term on the right-hand side of (4.28) appears because of the
dependence of $ \tilde v $ on $ f $ as is seen from (4.23). Inequality
(4.28) is proved in the same way as estimate (4.7.2) of \cite{[mb]}.

Combining (4.26)--(4.28) and using again the equality
$ |{\stackrel h{\nabla}}u|^2=|z|^2+|Hu|^2 $, we obtain
$$
\int\limits_{\Omega M}\Big[
(1-\beta-C\delta-C\varepsilon)|z|^2
+
(2-C\delta-C\varepsilon)|Hu|^2
-
\frac {4} {\beta}|\pi_\xi f^+\xi|^2
+
2|\bar f{}^-|^2
-4|f^-\xi|^2
-
C\varepsilon |f|^2\Big]d\Sigma
\leq
$$
$$
\leq
D\Big(
\|u|_{\partial_+\Omega M}\|^2_{H^1}+
\|f|_{\partial M}\|_{L^2}\cdot
\|u|_{\partial_+\Omega M}\|_{L^2}\Big).
                                                \eqno{(4.29)}
$$

The tensors $ f^+ $ and $ f^- $ are orthogonal to each other as well as
$ P_\xi f^+ $ and $ P_\xi f^- $ are orthogonal to each other. Therefore
$$
|f|^2=|f^+|^2+|f^-|^2,\quad
|P_\xi f|^2=|P_\xi f^+|^2+|P_\xi f^-|^2.
$$
With the help of these equalities, (3.45) gives
$$
|Hu|^2\geq |P_\xi f^+|^2-C\varepsilon|f^+|^2
-C\varepsilon|f^-|^2.
                        \eqno{(4.30)}
$$

We use (4.30) to transform the estimate (4.29) to the final form
$$
(1-\beta-C\delta-C\varepsilon)
\int\limits_{\Omega M}|z|^2 d\Sigma
+
\int\limits_{\Omega M}
\Big[(2-C\delta-C\varepsilon)|P_\xi f^+|^2
-
\frac {4} {\beta}|\pi_\xi f^+\xi|^2
-
C\varepsilon|f^+|^2
\Big]d\Sigma
+
$$
$$
+
2\int\limits_{\Omega M}
(|\bar f{}^-|^2-2|f^-\xi|^2-C\varepsilon|f^-|^2)d\Sigma
\leq
D\Big(
\|u|_{\partial_+\Omega M}\|^2_{H^1}+
\|f|_{\partial M}\|_{L^2}\cdot
\|u|_{\partial_+\Omega M}\|_{L^2}\Big)
                                                \eqno{(4.31)}
$$
with some new constant $ C $.

For $ \beta=1 $ and $ \varepsilon=\delta=0 $, the left-hand side of
(4.31) is the integral over $ M $ of the Hermitian form
$$
A(f,f)=B(f^+,f^+)+2Q(f^-,f^-),
$$
where $ B $ is the same as in Lemma 3.2 and
$$
Q(f,f)=\int\limits_{\Omega_xM}
(|f|^2-2|f\xi|^2)d\omega_x(\xi).
                        \eqno{(4.32)}
$$
By Lemma 3.2, the form $ B $ is positively definite on symmetric
tensors in the 3-dimensional case. The form $ Q $ is positively
definite on skew-symmetric tensors. Indeed, repeating the arguments of
the proof of Lemma 3.2, we derive in the 3-dimensional case
$$
\frac {1} {\omega}\int\limits_{\Omega_xM}
(|f|^2-2|f\xi|^2)d\omega_x(\xi)=
\frac {1} {3}|f|^2
$$
for a skew-symmetric tensor $ f $. Now, the proof is finished in the
same way as in Theorem 3.1.

\section[]{Nonlinear problem}

We return to considering the inverse problem of recovering a tensor
field $ f\in C^\infty(\tau^1_1M) $ on a CNTM $ (M,g) $ from the data
$$
\Phi[f]=U|_{\partial_+\Omega M},
                        \eqno{(5.1)}
$$
where $ U\in C(\beta^1_1M;\Omega M) $
is the solution to the boundary value problem on
$ \Omega M $
$$
HU=(P_\xi f)U,\quad \quad U|_{\partial_-\Omega M}=E.
                        \eqno{(5.2)}
$$
We will prove the uniqueness under the following smallness assumptions
on $ f $:
$$
|f(x)|<\varepsilon \quad \mbox{for}\quad x\in\partial M,
                        \eqno{(5.3)}
$$
$$
\int\limits_{\tau_-(x,\xi)}^{0}
|f(\gamma_{x,\xi}(t))|dt<\varepsilon,\quad
\int\limits_{\tau_-(x,\xi)}^{0}
|{\nabla}f(\gamma_{x,\xi}(t))|dt<\varepsilon\quad
\mbox{for}\quad (x,\xi)\in\partial_+\Omega M.
                        \eqno{(5.4)}
$$
We remind that we use notations introduced in Section 2. In particular,
$ \nabla $ is the covariant derivative.
Note that these smallness conditions are quite similar to that of
Theorem 2 of \cite{[V]}.

\begin{Theorem}
It is possible to choose a positive number $ \delta=\delta(n) $ for
$ n\geq 4 $ such that, for an $ n $-dimensional CNTM $ (M,g) $
satisfying the curvature condition (3.7), there exists a positive
number $ \varepsilon=\varepsilon(M,g) $ such that the following
statement is true. Let two tensor fields $ f_i\in C^\infty(\tau^1_1M) $
$ (i=1,2) $ satisfy (5.3)--(5.4) and $ \Phi_i=\Phi[f_i] $ be the
corresponding data. Then the estimate
$$
\|f_2-f_1\|_{L^2}\leq C\|\Phi^{-1}_1\Phi_2-E\|_{H^1}
                        \eqno{(5.5)}
$$
holds with a constant $ C $ independent of $ f_i $. In particular,
$ f_1=f_2 $ if $ \Phi_1=\Phi_2 $. In the case of $ n=3 $, the same
statement is true under the additional assumption that $ f_2-f_1 $ is
a closed tensor field.
\end{Theorem}

{\bf Proof.}
Let $ U_i\in C(\beta^1_1M;\Omega M)\ (i=1,2) $
be the solution to the boundary value problem
$$
HU_i=(P_\xi f_i)U_i,\quad \quad U_i|_{\partial_-\Omega M}=E.
                        \eqno{(5.6)}
$$
According to (5.2), the solution satisfies
$$
U_i\xi=U^*_i\xi=\xi .
                        \eqno{(5.7)}
$$
Set $ u=U^{-1}_1U_2-E $. Using the equalities $ HE=0 $,
$ H(U^{-1}_1U_2)=HU^{-1}_1\cdot U_2+U^{-1}_1HU_2 $, and
$ HU^{-1}_1=-U^{-1}_1HU_1\cdot U^{-1}_1 $, one easily derive from (5.6)
that $ u $ solves the boundary value problem
$$
Hu=U^{-1}_1\big(P_\xi(f_2-f_1)\big)U_2,\quad\quad
u|_{\partial_-\Omega M}=0
                        \eqno{(5.8)}
$$
and
$$
u|_{\partial_+\Omega M}=\Phi^{-1}_1\Phi_2-E.
                        \eqno{(5.9)}
$$
Setting $ f=f_2-f_1,\ p=U^{-1}_1,\ q=U_2 $, we write (5.8)--(5.9) in
the form
$$
Hu=p(P_\xi f)q,\quad \quad u|_{\partial_-\Omega M}=0,
                        \eqno{(5.10)}
$$
$$
F[f]=u|_{\partial_+\Omega M}=\Phi^{-1}_1\Phi_2-E.
                        \eqno{(5.11)}
$$
We have arrived to the linear problem considered in Sections 3 and 4.
If we will prove that the weights $ p=U^{-1}_1 $ and $ q=U_2 $ satisfy
conditions (3.1) and (3.5), we would be able to apply Theorems 3.1 and
4.2 to obtain the statement of Theorem 5.1.
Condition (3.1) is satisfied by (5.7).
The weight $ p=U^{-1}_1 $
solves the boundary value problem
$$
Hp=-p(P_\xi f_1),\quad \quad p|_{\partial_-\Omega M}=E
$$
which is very similar to (5.2). Theorem 5.1 is thus reduced to the
following

\begin{Lemma}
Let $ (M,g) $ be a CNTM and a tensor field $ f\in C^\infty(\tau^1_1M)
$ satisfy (5.3)--(5.4). Then the solution $ U $ to the boundary value
problem (5.2) satisfies the estimates
$$
|U-E|<C\varepsilon,\quad
|{\stackrel v{\nabla}}U|<C\varepsilon\quad
\mbox{on}\quad \Omega M
                        \eqno{(5.12)}
$$
with some constant $ C $ depending on $ (M,g) $ but not on $ f $.
\end{Lemma}

To prove Lemma 5.2, we need the following estimate for solutions to
linear ordinary differential equations (see Lemma 4.1 of Chapter IV
of \cite{[H]}):

\begin{Lemma}
Let $ y=(y_1(t),\dots,y_N(t)) $ be the solution to the initial value
problem
$$
\frac {dy} {dt}=A(t)y+f(t),\quad y(0)=y_0,
$$
where $ f(t) $ is an $ N $-dimensional vector and $ A(t) $ is an
$ N\times N $-matrix. Then
$$
|y(t)|\leq\Big(|y_0|+
\int\limits_{0}^{t}|f(\tau)|d\tau\Big)
\exp\int\limits_{0}^{t}|A(\tau)|d\tau,
$$
where $ |A(\tau)| $ is the operator norm of the matrix $ A(\tau) $
defined with the help of the standard norm $ |\cdot| $ on
$ {\mathbb C}^N $.
\end{Lemma}

Let us adjust this statement to our geometric setting.

\begin{Lemma}
Let $ (M,g) $ be a CNTM and $ f\in C^\infty(\beta^0_mM;\Omega M) $,
$ A\in C^\infty(\beta^m_mM;\Omega M) $ be two semibasic tensor fields.
Let $ y\in C(\beta^0_mM;\Omega M) $ be the solution to the
boundary value problem
$$
Hy=Ay+f,\quad \quad y|_{\partial_-\Omega M}=y_0
$$
with some $ y_0\in C(\beta^0_mM;{\partial_-\Omega M}) $. Then
$$
|y(x,\xi)|\leq
C\Big(|y_0(\gamma(\tau_-(x,\xi)),\dot\gamma(\tau_-(x,\xi)))|
+\!\!\!\int\limits_{\tau_-(x,\xi)}^{0}
|f(\gamma(t),\dot\gamma(t))|dt\Big)\times
$$
$$
\times
\exp\Big[C\!\!\!\int\limits_{\tau_-(x,\xi)}^{0}
|A(\gamma(t),\dot\gamma(t))|dt\Big]
$$
for $ (x,\xi)\in\Omega M $, where $ \gamma=\gamma_{x,\xi} $.
Here the norm $ |\cdot| $ is defined in Section 2, and the constant
$ C $ depends on $ m $ and $ (M,g) $.
\end{Lemma}

The constant $ C $ appears in Lemma 5.4 since different norms are used
in this Lemma and Lemma 5.3. In what follows in this section, we denote
different constants depending on
$ (M,g) $ by the same letter $ C $.

{\bf Proof of Lemma 5.2.}
Let $ U $ be the solution to (5.2). Then $ U-E $ solves the boundary
value problem
$$
H(U-E)=(P_\xi f)(U-E)+P_\xi f,\quad (U-E)|_{\partial_-\Omega M}=0.
$$
Applying Lemma 5.4, we obtain the estimate
$$
|(U-E)(x,\xi)|\leq
C\!\!\!\int\limits_{\tau_-(x,\xi)}^{0}
\!\!\!|P_{\dot\gamma(t)}f(\gamma(t))|dt\
\exp\Big[ C\!\!\!\int\limits_{\tau_-(x,\xi)}^{0}
\!\!\!|P_{\dot\gamma(t)}f(\gamma(t))|dt\Big]
\leq
$$
$$
\leq
C\!\!\!\int\limits_{\tau_-(x,\xi)}^{0}
\!\!\!|f(\gamma(t))|dt
\exp\Big[C\!\!\!\int\limits_{\tau_-(x,\xi)}^{0}
\!\!\!|f(\gamma(t))|dt\Big],
$$
Together with (5.4),
this implies the first of inequalities (5.12).

The proof of the second of estimates (5.12) is more troublesome
because $ {\stackrel v{\nabla}}U $ and $ {\stackrel h{\nabla}}U $
must be estimated together but $ {\stackrel h{\nabla}}U $ is unbounded
near $ \Omega(\partial M) $.

We start with estimating $ {\stackrel h{\nabla}}U $ on
$ \partial_-\Omega M $. To this end we consider equation
(5.2) at
a boundary point $ (x,\xi)\in \partial_-\Omega M $. Because of the
boundary condition $ U|_{\partial_-\Omega M}=E $, equation (5.2) gives
$$
(HU)|_{\partial_-\Omega M}=P_\xi f.
                        \eqno{(5.13)}
$$
Let us choose boundary normal coordinates $ (x^1,\dots,x^n) $ in a
neighborhood $ V $
of the boundary point such that the boundary is determined
by the equation $ x^n=0 $, $ x^n\geq 0 $ in $ V $,
and $ g_{in}=\delta_{in} $. Because of the
boundary condition $ U^i_j|_{x^n=0}=\delta^i_j $, the equalities
$$
{\stackrel h{\nabla}}_{\!\alpha}U^i_j|_{x^n=0}=0
\quad (0\leq\alpha \leq n-1)
$$
hold, and equation (5.13) becomes
$$
(\xi_n{\stackrel h{\nabla}}_{\!n}U^i_j)|_{x^n=0}=
(P_\xi f)^i_j.
$$
Since $ \xi_n=-\langle \xi,\nu\rangle $,
where $ \nu=\nu(x) $ is the unit outward normal to the boundary,
this gives with the help of
(5.3)
$$
|{\stackrel h{\nabla}}U(x,\xi)|\leq
\frac {C\varepsilon} {|\langle\xi,\nu\rangle|}\quad
\mbox{for}\quad (x,\xi)\in \partial_-\Omega M,\
\langle\nu,\xi\rangle\neq 0.
                        \eqno{(5.14)}
$$

$ {\stackrel v{\nabla}}U $ vanishes on $ \partial_-\Omega M $ as
follows from the condition $ U|_{\partial_-\Omega M}=E $,
$$
{\stackrel v{\nabla}}U|_{\partial_-\Omega M}=0.
                        \eqno{(5.15)}
$$

Applying the operators $ {\stackrel v{\nabla}} $ and
$ {\stackrel h{\nabla}} $ to equation (5.2), we obtain
$$
{\stackrel v{\nabla}}HU=(P_\xi f){\stackrel v{\nabla}}U
+({\stackrel v{\nabla}}P_\xi) fU,\quad
{\stackrel h{\nabla}}HU=(P_\xi f){\stackrel h{\nabla}}U
+(P_\xi{\nabla}f)U.
                        \eqno{(5.16)}
$$
We have used that $ {\stackrel v{\nabla}}f=0 $ and
$ {\stackrel h{\nabla}}f={\nabla}f $ since $ f $ is independent of
$ \xi $.

Operators $ {\stackrel v{\nabla}} $ and $ H $ satisfy the commutator
formula
$$
{\stackrel v{\nabla}}H=H{\stackrel v{\nabla}}
+{\stackrel h{\nabla}}
                        \eqno{(5.17)}
$$
as follows immediately from the definition
$ H=\xi^i{\stackrel h{\nabla}}_{\!i} $ and from the fact that
$ {\stackrel v{\nabla}} $ and $ {\stackrel h{\nabla}} $ commute.
The commutator formula for $ {\stackrel h{\nabla}} $ and $ H $ is a
little bit more complicated. Indeed, using the commutator formula for
horizontal derivatives (Theorem 3.5.2 of \cite{[mb]}), we see that
$$
{\stackrel h{\nabla}}_{\!i}(HU)^j_k
=
{\stackrel h{\nabla}}_{\!i}
(\xi^p{\stackrel h{\nabla}}_{\!p}U^j_k)
=
\xi^p{\stackrel h{\nabla}}_{\!i}{\stackrel h{\nabla}}_{\!p}U^j_k
=
$$
$$
=
\xi^p{\stackrel h{\nabla}}_{\!p}
{\stackrel h{\nabla}}_{\!i}U^j_k
+
\xi^p\Big(
-R^l_{\,qip}\xi^q
{\stackrel v{\nabla}}_{\!l}U^j_k
+
R^j_{\,lip}U^l_k-R^l_{\,kip}U^j_l\Big).
$$
This can be written as
$$
{\stackrel h{\nabla}}HU=H{\stackrel h{\nabla}}U
+{\cal R}_1[{\stackrel v{\nabla}}U]+
{\cal R}_2[U]
                        \eqno{(5.18)}
$$
with some algebraic operators $ {\cal R}_1 $ and $ {\cal R}_2 $ on
semibasic tensors which are determined by the curvature tensor. The
operator $ {\cal R}_2 $ satisfies $ {\cal R}_2[E]=0 $. Therefore the
first of estimates (5.12), which is already proved, implies the
inequality
$$ |{\cal R}_2[U]|=|{\cal R}_2[U-E]|\leq C\varepsilon
                        \eqno{(5.19)}
$$
with some constant $ C $ depending on the curvature bound.

Using commutator formulas (5.17) and (5.18), we write (5.16) as
$$
H({\stackrel v{\nabla}}U)=
(P_\xi f){\stackrel v{\nabla}}U-
{\stackrel h{\nabla}}U+F,
                        \eqno{(5.20)}
$$
$$
H({\stackrel h{\nabla}}U)=
-{\cal R}_1[{\stackrel v{\nabla}}U]+
(P_\xi f){\stackrel h{\nabla}}U+G,
                        \eqno{(5.21)}
$$
where
$$
F=({\stackrel v{\nabla}}P_\xi)fU,\quad
G=(P_\xi{\nabla}f)U-{\cal R}_2[U].
                        \eqno{(5.22)}
$$

We first consider (5.20)--(5.21) as a linear system of ordinary
differential equations in coordinates of $ {\stackrel v{\nabla}}U $ and
$ {\stackrel h{\nabla}}U $ with free terms $ F $
and $ G $. Applying Lemma 5.4
to the system, we obtain the estimate
$$
|{\stackrel h{\nabla}}U(x,\xi)|
\leq
C\Big\{
|{\stackrel v{\nabla}}U(\gamma(\tau_-(x,\xi)),
\dot\gamma(\tau_-(x,\xi)))|
+
|{\stackrel h{\nabla}}U(\gamma(\tau_-(x,\xi)),
\dot\gamma(\tau_-(x,\xi)))|
+
$$
$$
+\!\!\!\!
\int\limits_{\tau_-(x,\xi)}^{0}\!\!\!\!
(|F(\gamma(t),\dot\gamma(t))|\!+\!|G(\gamma(t),\dot\gamma(t))|
)dt
\Big\}
\exp\Big[C\!\!\!\!\int\limits_{\tau_-(x,\xi)}^{0}
\!\!\!\!(
|{\cal R}_1((\gamma(t),\dot\gamma(t))|
\!+\!
|P_{\dot\gamma(t)}f(\gamma(t))|\!+\!|E|)dt\Big].
                        \eqno{(5.23)}
$$
The first summand of the expression in braces is equal to zero by (5.15). In virtue of
(5.14), the second summand of this expression is estimated as
$$
|{\stackrel h{\nabla}}U(\gamma(\tau_-(x,\xi)),
\dot\gamma(\tau_-(x,\xi)))|
\leq
\frac {C\varepsilon}
{|\langle\dot\gamma(\tau_-(x,\xi)),
\nu(\gamma(\tau_-(x,\xi))\rangle|}.
$$
By Lemma 4.1.2 of \cite{[mb]}, the estimate
$$
\frac {|\tau_-(x,\xi)|}
{|\langle\dot\gamma(\tau_-(x,\xi)),\nu(\gamma(\tau_-(x,\xi)))\rangle|}
\leq C
$$
holds.
Combining two last estimates, we obtain
$$
|{\stackrel h{\nabla}}U(\gamma(\tau_-(x,\xi)),
\dot\gamma(\tau_-(x,\xi)))|
\leq
\frac {C\varepsilon}
{|\tau_-(x,\xi)|}.
$$
As is seen from (5.22), (5.19), and
(5.4), the integral inside the braces in (5.23)
can be estimated by $ C\varepsilon $.
Finally, the integral under the exponent on (5.23) is estimated by some
constant. Therefore (5.23) implies the estimate
$$
|{\stackrel h{\nabla}}U(x,\xi)|\leq
\frac {C\varepsilon} {|\tau_-(x,\xi)|}\quad
\mbox{for}\quad (x,\xi)\in\Omega M.
                        \eqno{(5.24)}
$$

Now, we consider (5.20) as a linear system of ordinary
differential equations in coordinates of $ {\stackrel v{\nabla}}U $
with the free term $ -{\stackrel h{\nabla}}U+F $. Applying Lemma 5.4
to the system and using the homogeneous initial condition (5.15), we
obtain the estimate
$$
|{\stackrel v{\nabla}}U(x,\xi)|\leq
C\!\!\!\int\limits_{\tau_-(x,\xi)}^{0}
\!\!\!\Big(|{\stackrel h{\nabla}}U(\gamma(t),\dot\gamma(t))|
+|F(\gamma(t),\dot\gamma(t))|\Big)dt
\exp\Big[C\!\!\!\int\limits_{\tau_-(x,\xi)}^{0}
\!\!\!|P_{\dot\gamma(t)}f(\gamma(t))|dt\Big].
                        \eqno{(5.25)}
$$
In virtue of (5.24), the first integral on the right-hand side of
(5.25) can be
estimated by $ C\varepsilon $. Estimating then the second integral by
(5.4), we obtain the second of inequalities (5.12). The lemma is
proved.

\section[]{Kernel of the operator $ S $}

In this section, we restrict ourselves to considering symmetric matrix
functions on the whole of $ {\mathbb R}^3 $ endowed with the standard
scalar product $ \langle\cdot,\cdot\rangle $.

Let $ M(3) $ be the space of complex-valued symmetric $ 3\times 3
$-matrices. Such a matrix $ f\in M(3) $ is considered as the linear
operator $ f:{\mathbb C}^3\rightarrow{\mathbb C}^3 $. By
$ {\mathbb S}^2 $ we denote the unit sphere in $ {\mathbb R}^3 $ and by
$ T{\mathbb S}^2=\{(x,\xi)\mid\xi\in{\mathbb S}^2,
x\in{\mathbb R}^3,\langle x,\xi\rangle=0\} $, the tangent bundle of the
sphere. Given $ \xi\in{\mathbb S}^2 $, let
$ \xi^\bot=\{\eta\in{\mathbb C}^3\mid\langle\eta,\xi\rangle=0\} $ be
the complex two-dimensional space of vectors orthogonal to $ \xi $ and
$ \pi_\xi:{\mathbb C}^3\rightarrow{\mathbb C}^3 $, the orthogonal
projection onto $ \xi^\bot $.

Let $ {\cal S}({\mathbb R}^3;M(3)) $ be the Schwartz space of
$ M(3) $-valued functions on $ {\mathbb R}^3 $. The linear operator
$$
S:{\cal S}({\mathbb R}^3;M(3))\rightarrow C^\infty(T{\mathbb S}^2)
$$
is defined by
$$
S[f](x,\xi)=\int\limits_{-\infty}^{\infty}
\mbox{tr}\,(\pi_\xi f(x+t\xi)\pi_\xi)dt\quad
\mbox{for}\quad (x,\xi)\in T{\mathbb S}^2,
                        \eqno{(6.1)}
$$
compare with (1.10). We are going to answer the question: to which
extent is a symmetric matrix function
$ f\in{\cal S}({\mathbb R}^3;M(3)) $
determined by the data $ S[f] $? Since $ S[f] $ depends
linearly on $ f $, the question is equivalent to studying the kernel of
the operator $ S $.

Let us recall the definition of the ray transform
$$
I:{\cal S}({\mathbb R}^3;M(3))\rightarrow C^\infty(T{\mathbb S}^2),
$$
$$
I[f](x,\xi)=\int\limits_{-\infty}^{\infty}
\langle f(x+t\xi)\xi,\xi\rangle dt=
\int\limits_{-\infty}^{\infty}
f_{ij}(x+t\xi)\xi^i\xi^j\,dt\quad
\mbox{for}\quad (x,\xi)\in T{\mathbb S}^2.
                        \eqno{(6.2)}
$$
See Chapter 2 of \cite{[mb]} for the theory of the ray transform on the
Euclidean space. Our approach to studying the kernel of $ S $ is
based on the following observation: $ S[f]=-I[f] $ for a symmetric matrix
function $ f $ with zero trace, see equation (6.7) below.

\begin{Theorem}
A symmetric matrix function
$ f=(f_{ij}(x))\in\mathcal{S}({\mathbb{R}}^3;M(3)) $
belongs to the kernel of the operator $ S $ if and only if
it satisfies the system
of partial differential equations
$$
\left.
\begin{array}{l}
R_1[f]:=f_{11;11}+2f_{12;12}+f_{22;22}+f_{33;11}+f_{33;22}=0\\
[0.3cm]
R_2[f]:=f_{11;11}+2f_{13;13}+f_{22;11}+f_{22;33}+f_{33;33}=0\\
[0.3cm]
R_3[f]:=f_{11;22}+f_{11;33}+2f_{23;23}+f_{22;22}+f_{33;33}=0
\end{array}\right\}
                                           \eqno{(6.3)}
$$
where the tensor notations for partial derivatives
$ f_{ij;kl}=\partial^2f_{ij}/\partial x_k\partial x_l $
are used for brevity.
\end{Theorem}

The theorem gives the full system
$ \{R_i[f]\mid i=1,2,3\} $
of local linear functionals that can be recovered from the data
$ S[f] $.

{\bf Proof.}
Represent the matrix $ f $ as
$$
f=\tilde f+\lambda E,\quad\quad
\mbox{tr}\,\tilde f=0,
                                     \eqno{(6.4)}
$$
where $ E $ is the unit matrix and $ \lambda\in\mathcal{S}({\mathbb{R}}^3) $
is a scalar function. Then
$$
S[f]=S[\tilde f]+S[\lambda E].
                                     \eqno{(6.5)}
$$

As follows from definition (6.1) of the operator $ S $,
$$
S[\lambda E]=2I[\lambda E],
                                     \eqno{(6.6)}
$$
where
$$
I[\lambda E](x,\xi)=
\int\limits_{-\infty}^{\infty}\lambda(x+t\xi)dt\quad
\mbox{for}\quad (x,\xi)\in T{\mathbb{S}}^2
$$
is the ray transform of the matrix function $ \lambda E $.

If $ (\xi,\eta,\zeta) $ is an orthonormal basis of
$ {\mathbb{R}}^3 $, then
$$
0=\mbox{tr}\,\tilde f=
\langle\tilde f(x)\xi,\xi\rangle+
\langle\tilde f(x)\eta,\eta\rangle+
\langle\tilde f(x)\zeta,\zeta\rangle
$$
and
$$
\mbox{tr}\,(\pi_\xi\tilde f(x)\pi_\xi)=
\langle\tilde f(x)\eta,\eta\rangle+
\langle\tilde f(x)\zeta,\zeta\rangle.
$$
This implies
$$
\mbox{tr}\,(\pi_\xi \tilde f(x+t\xi)\pi_\xi)=
-\langle\tilde f(x+t\xi)\xi,\xi\rangle.
$$
Integrating the last equality with respect to $ t $ and recalling
definitions (6.1) and (6.2) of $ S $ and $ I $, we see that
$$
S[\tilde f]=-I[\tilde f].
                                     \eqno{(6.7)}
$$
Substituting (6.6)--(6.7) into (6.5), we obtain
$$
S[f]=-I[\tilde f-2\lambda E].
$$
Thus, $ f $ is in the kernel of $ S $ if and only if
$$
I[\tilde f-2\lambda E]=0.
                                   \eqno{(6.8)}
$$

Now, we apply Theorem 2.2.1 of \cite{[mb]} which states that (6.8)
is equivalent to the existence of a vector field $ v $
such that
$$
\tilde f-2\lambda E=dv,
                                   \eqno{(6.9)}
$$
where
$$
d:C^\infty({\mathbb{R}}^3;{\mathbb{R}}^3)\rightarrow
C^\infty({\mathbb{R}}^3;M(3)),
\quad
(dv)_{ij}=\frac{1}{2}\Big(
\frac{\partial v_i}{\partial x_j}+
\frac{\partial v_j}{\partial x_i}\Big)
$$
is the inner derivative. Theorem 2.2.1 of \cite{[mb]} is formulated
and proved for compactly
supported tensor fields. Nevertheless, the same proof works for
$ \tilde f-2\lambda E\in \mathcal{S}({\mathbb{R}}^3;M(3)) $
and gives the vector field $ v $ belonging to the Schwartz space
$ \mathcal{S}({\mathbb{R}}^3;{\mathbb{R}}^3) $.

Let us express $ \lambda $ and $ \tilde f $ through $ f $. Applying the
trace operator to the first of equations (6.4) and taking the second one into
account, we see that
$$
\lambda=\frac {1} {3}\mbox{tr}\,f.
                        \eqno{(6.10)}
$$
From (6.4) and (6.10),
$$
{\tilde f}_{ij}=f_{ij}-\frac {1} {3}\mbox{tr}\,f\cdot\delta_{ij}.
                        \eqno{(6.11)}
$$
Substitute (6.10)--(6.11) into (6.9) to obtain
$$
dv=f-\mbox{tr}\,f\cdot E.
                        \eqno{(6.12)}
$$

Equation (6.12) represents the overdetermined system of six first order
partial differential equations in three unknowns $ (v_1,v_2,v_3) $. The
solvability condition for the system  is presented by Theorem 2.2.2 of
\cite{[mb]}:
equation (6.12) is solvable if and only if the right-hand side of
(6.12) belongs to the kernel of the Saint-Venant operator. Here, we
prefer to use the version $ R $ of the Saint-Venant operator which is
defined by the equation before formula (2.4.6) of
\cite{[mb]}. So, the
solvability condition for (6.12) is
$$
Rh=0,\quad \mbox{where}\quad h=f-\mbox{tr}\,f\cdot E.
                        \eqno{(6.13)}
$$

The operator $ R $ is defined by the formula
$$
4(Rh)_{ijkl}=h_{ik;jl}-h_{jk;il}-h_{il;jk}+h_{jl;ik}.
$$
It possesses the following symmetries:
$$
(Rh)_{ijkl}=-(Rh)_{jikl}=-(Rh)_{ijlk}=(Rh)_{klij}.
$$
Because of the symmetries, the tensor $ Rh $ has six linearly
independent components, and equation (6.13) is equivalent to the system
$$
\begin{array}{l}
-R_1[f]:=(Rh)_{1212}=h_{11;22}-2h_{12;12}+h_{22;11}=0,\\
[0.2cm]
-R_2[f]:=(Rh)_{1313}=h_{11;33}-2h_{13;13}+h_{33;11}=0,\\
[0.2cm]
-R_3[f]:=(Rh)_{2323}=h_{22;33}-2h_{23;23}+h_{33;22}=0,\\
[0.2cm]
-R_4[f]:=(Rh)_{1213}=h_{11;23}-h_{12;13}-h_{13;12}+h_{23;11}=0,\\
[0.2cm]
-R_5[f]:=(Rh)_{2123}=h_{22;13}-h_{12;23}-h_{23;12}+h_{13;22}=0,\\
[0.2cm]
-R_6[f]:=(Rh)_{1323}=h_{33;12}-h_{13;23}-h_{23;13}+h_{12;33}=0.
\end{array}
$$
Substitute the value
$ h_{ij}=f_{ij}-(f_{11}+f_{22}+f_{33})\delta_{ij} $ into the last system
$$
\left.
\begin{array}{l}
R_1[f]:=f_{11;11}+2f_{12;12}+f_{22;22}+f_{33;11}+f_{33;22}=0,\\
[0.2cm]
R_2[f]:=f_{11;11}+2f_{13;13}+f_{22;11}+f_{22;33}+f_{33;33}=0,\\
[0.2cm]
R_3[f]:=f_{11;22}+f_{11;33}+2f_{23;23}+f_{22;22}+f_{33;33}=0,\\
[0.2cm]
R_4[f]:=f_{12;13}+f_{13;12}+f_{22;23}-f_{23;11}+f_{33;23}=0,\\
[0.2cm]
R_5[f]:=f_{11;13}+f_{12;23}-f_{13;22}+f_{23;12}+f_{33;13}=0,\\
[0.2cm]
R_6[f]:=f_{11;12}-f_{12;33}+f_{13;23}+f_{22;12}+f_{23;13}=0.
\end{array}
\right\}
                        \eqno{(6.14)}
$$

For a symmetric matrix function
$ f=(f_{ij}(x))\in\mathcal{S}({\mathbb{R}}^3;M(3)) $,
system (6.14) contains only three independent equations. More
precisely: each of the last three equations of (6.14) can be obtained
from the first three equations by taking linear combinations,
differentiation, and integration. To prove this, let us rewrite system
(6.14) in terms of the Fourier transform $ g(\xi)=\hat f $. Applying the
Fourier transform to each equation of (6.14), we arrive to the system
$$
\left.
\begin{array}{l}
{\hat R}_1[g]:=
\xi_1^2g_{11}+2\xi_1\xi_2g_{12}+\xi_2^2g_{22}+
(\xi_1^2+\xi_2^2)g_{33}=0,\\
[0.2cm]
{\hat R}_2[g]:=
\xi_1^2g_{11}+2\xi_1\xi_3g_{13}+(\xi_1^2+\xi_3^2)g_{22}+
\xi_3^2g_{33}=0,\\
[0.2cm]
{\hat R}_3[g]:=
(\xi_2^2+\xi_3^2)g_{11}+\xi_2^2g_{22}+2\xi_2\xi_3g_{23}+
\xi_3^2g_{33}=0,\\
[0.2cm]
{\hat R}_4[g]:=
\xi_1\xi_3g_{12}+\xi_1\xi_2g_{13}+\xi_2\xi_3g_{22}-
\xi_1^2g_{23}+\xi_2\xi_3g_{33}=0,\\
[0.2cm]
{\hat R}_5[g]:=
\xi_1\xi_3g_{11}+\xi_2\xi_3g_{12}-\xi_2^2g_{13}+
\xi_1\xi_2g_{23}+\xi_1\xi_3g_{33}=0,\\
[0.2cm]
{\hat R}_6[g]:=
\xi_1\xi_2g_{11}-\xi_3^2g_{12}+\xi_2\xi_3g_{13}+
\xi_1\xi_2g_{22}+\xi_1\xi_3g_{23}=0.
\end{array}
\right\}
                        \eqno{(6.15)}
$$

One can easily see the following three relations between equations of
system (6.15):
$$
2\xi_2\xi_3 {\hat R}_4[g]=
\xi_3^2{\hat R}_1[g]+\xi_2^2{\hat R}_2[g]-\xi_1^2{\hat R}_3[g],
$$
$$
2\xi_1\xi_3 {\hat R}_5[g]=
\xi_3^2{\hat R}_1[g]-\xi_2^2{\hat R}_2[g]+\xi_1^2{\hat R}_3[g],
$$
$$
2\xi_1\xi_2 {\hat R}_5[g]=
-\xi_3^2{\hat R}_1[g]+\xi_2^2{\hat R}_2[g]+\xi_1^2{\hat R}_3[g].
$$
Therefore three last equations of (6.15) follow from three first
equations at least if $ g(\xi) $ depends continuously on $ \xi $.

Deleting three last equations from system (6.15), we obtain the
equivalent system
$$
\left.
\begin{array}{l}
{\hat R}_1[g]:=
\xi_1^2g_{11}+2\xi_1\xi_2g_{12}+\xi_2^2g_{22}+
(\xi_1^2+\xi_2^2)g_{33}=0,\\
[0.2cm]
{\hat R}_2[g]:=
\xi_1^2g_{11}+2\xi_1\xi_3g_{13}+(\xi_1^2+\xi_3^2)g_{22}+
\xi_3^2g_{33}=0,\\
[0.2cm]
{\hat R}_3[g]:=
(\xi_2^2+\xi_3^2)g_{11}+\xi_2^2g_{22}+2\xi_2\xi_3g_{23}+
\xi_3^2g_{33}=0.
\end{array}
\right\}
                        \eqno{(6.16)}
$$
The same is true for system (6.14): deleting three last equations from
(6.14), we will obtain the equivalent system (6.3). The theorem is
proved.

\bigskip

System (6.16) enables us to answer the question: which integral moments of
$ f $ can be determined from $ S[f] $? Indeed,
the Tailor series of the
function $ g(\xi)=\hat f $ is
$$
g_{jk}(\xi)\sim
\sum\limits_{m=0}^{\infty}\sum\limits_{|\alpha|=m}
\frac {i^m } {\alpha!}\mu^{(m)}_{jk,\alpha}[f]\xi^\alpha,
                        \eqno{(6.17)}
$$
where
$$
\mu^{(m)}_{jk,\alpha}[f]=
\int\limits_{{\mathbb R}^3}x^\alpha f_{jk}(x)dx
\quad \quad |\alpha|=m
$$
are the integral moments of order $ m $.
Assuming $ f $ to be in the kernel of $ S $, let us insert series
(6.17) into system (6.16). Since the coefficients of (6.16) are
homogeneous functions of $ \xi $, the system does not mix moments
of different orders. This means that we can take $ g(\xi) $ in the form
$$
g_{jk}(\xi)=
\sum\limits_{|\alpha|=m}
\frac {i^m } {\alpha!}\mu^{(m)}_{jk,\alpha}[f]\xi^\alpha
$$
if we are looking for moments of order $ m $.

Let us start with considering zero order moments. We substitute the
expressions
$
g_{ij}=
\mu^{(0)}_{ij}
$
into (6.16). Equating coefficients at the same degrees of
$ \xi $ at the resulting equations, we easily find that
$
\mu^{(0)}_{ij}=0
$ for every $ (i,j) $. This means that
the integral
$ \int_{{\mathbb R}^3}f(x)dx $ can be determined from the data
$ S[f] $.

Next, we consider first order moments. We substitute the
expressions
$$
g_{ij}(\xi)=
\mu^{(1)}_{ij,1}\xi_1+\mu^{(1)}_{ij,2}\xi_2+\mu^{(1)}_{ij,3}\xi_3
$$
into system (6.16). Equating coefficients at the same degrees of
$ \xi $ at the resulting equations, we arrive to the system
$$
\mu^{(1)}_{11,1}+\mu^{(1)}_{22,1}=0,\quad
\mu^{(1)}_{11,1}+\mu^{(1)}_{33,1}=0,\quad
\mu^{(1)}_{11,2}+\mu^{(1)}_{22,2}=0,
$$
$$
\mu^{(1)}_{22,2}+\mu^{(1)}_{33,2}=0,\quad
\mu^{(1)}_{11,3}+\mu^{(1)}_{33,3}=0,\quad
\mu^{(1)}_{22,3}+\mu^{(1)}_{33,3}=0,
$$
$$
\mu^{(1)}_{11,2}+2\mu^{(1)}_{12,1}+\mu^{(1)}_{33,2}=0,\quad
\mu^{(1)}_{11,3}+2\mu^{(1)}_{13,1}+\mu^{(1)}_{22,3}=0,\quad
\mu^{(1)}_{11,3}+\mu^{(1)}_{22,3}+2\mu^{(1)}_{23,2}=0,
$$
$$
2\mu^{(1)}_{12,2}+\mu^{(1)}_{22,1}+\mu^{(1)}_{33,1}=0,\quad
2\mu^{(1)}_{13,3}+\mu^{(1)}_{22,1}+\mu^{(1)}_{33,1}=0,\quad
\mu^{(1)}_{11,2}+2\mu^{(1)}_{23,3}+\mu^{(1)}_{33,2}=0.
$$
The general solution to the system looks as follows:
$$
\mu^{(1)}_{11,1}=a_1,\quad
\mu^{(1)}_{12,1}=a_2,\quad
\mu^{(1)}_{13,1}=a_3,\quad
\mu^{(1)}_{22,1}=-a_1,\quad
\mu^{(1)}_{23,1}=0,\quad
\mu^{(1)}_{33,1}=-a_1,
$$
$$
\mu^{(1)}_{11,2}=-a_2,\quad
\mu^{(1)}_{12,2}=a_1,\quad
\mu^{(1)}_{13,2}=0,\quad
\mu^{(1)}_{22,2}=a_2,\quad
\mu^{(1)}_{23,2}=a_3,\quad
\mu^{(1)}_{33,2}=-a_2,
$$
$$
\mu^{(1)}_{11,3}=-a_3,\quad
\mu^{(1)}_{12,3}=0,\quad
\mu^{(1)}_{13,3}=a_1,\quad
\mu^{(1)}_{22,3}=-a_3,\quad
\mu^{(1)}_{23,3}=a_2,\quad
\mu^{(1)}_{33,3}=a_3,
$$
where $ (a_1,a_2,a_3) $ are arbitrary constants. Eliminating the
constants, we obtain the following independent system of 15 linear
combinations of first order moments of $ f $ which can be recovered
from the data $ S[f] $:
$$
(\mu^{(1)}_{ij,k}+\delta_{ij}\mu^{(1)}_{kk,k}-
\delta_{ik}\mu^{(1)}_{jj,j}-\delta_{jk}\mu^{(1)}_{ii,i})[f],
                        \eqno{(6.18)}
$$
where $ \delta_{ij} $ is the Kronecker tensor.
System (6.18) is considered for
such $ (i,j,k) $ that at least two of these indices are different.
A similar consideration is possible for integral moments
$ \mu^{(m)}_{jk,\alpha}[f] $ of an arbitrary order $ m $.

\end{document}